\documentclass[%
 reprint,
superscriptaddress,
 amsmath,amssymb,
aps,
pra,
floatfix,
citeautoscript
]{revtex4-2}

\usepackage{tikz}
\usepackage{graphicx}
\usepackage{dcolumn}
\usepackage{bm}
\usepackage{graphicx}
\usepackage{pgfplots}
\usepackage{float}
\usepackage[colorlinks=true,   
            linkcolor=blue,    
            citecolor=blue,     
            urlcolor=blue,     
            pdfusetitle]       
            {hyperref}

\begin{document}

\preprint{APS/123-QED}

\title{Simultaneous Determination of Local Magnetic Fields and Sensor Orientation with Nitrogen-Vacancy Centers in Nanodiamond}

\author{Yizhou Wang}
\thanks{These authors contributed equally to this work.}
\author{Haochen Shen}
\thanks{These authors contributed equally to this work.}
\author{Zhongyuan Liu}
\author{Yue Yu}
\affiliation{Department of Physics, Washington University in St. Louis, St. Louis, MO, USA}

\author{Shengwang Du}

\affiliation{Elmore Family School of Electrical and Computer Engineering,
Purdue University, West Lafayette, IN, USA}
\affiliation{Department of Physics and Astronomy, Purdue University, West Lafayette, IN, USA}
\affiliation{Purdue Quantum Science and Engineering Institute,
Purdue University, West Lafayette, IN, USA}
\author{Chong Zu}
\email{zu@wustl.edu}
\affiliation{Department of Physics, Washington University in St. Louis, St. Louis, MO, USA}
\affiliation{Center for Quantum Leaps, Washington University, St. Louis, MO, USA}

\author{Chuanwei Zhang}
\email{chuanwei.zhang@wustl.edu}
\affiliation{Department of Physics, Washington University in St. Louis, St. Louis, MO, USA}
\affiliation{Center for Quantum Leaps, Washington University, St. Louis, MO, USA}

\date{\today}

\begin{abstract}
Nitrogen-vacancy (NV) centers in nanodiamonds have emerged as a promising quantum sensing platform for biomedical imaging applications, yet random orientations of individual particles present significant challenges in large-scale sensor calibration.
In this study, we demonstrate a novel approach to simultaneously determine each particle's crystallographic axes and the surrounding local vector magnetic field.
Specifically, a minimum of four distinct bias fields is required to unambiguously extract both the orientation and the local field.
We validate our method experimentally using NV centers in two scenarios: (1) in a bulk diamond with known crystal orientation as a proof of concept, and (2) on various single nanodiamonds to mimic real-world applications.
Our work represents a crucial step towards unlocking the full potential of nanodiamonds for advanced applications such as in-situ biomedical imaging and nanoscale sensing in complex environments.

 \end{abstract}

\maketitle

\section{Introduction}

Quantum sensing with nitrogen-vacancy (NV) centers in diamond has enabled precise measurements of magnetic~\cite{Balasubramanian2008, Maze2008, PhysRevB.80.115202, PhysRevLett.106.030802, PhysRevApplied.22.044030, RevModPhys.92.015004,PhysRevB.80.115202,  PhysRevLett.108.197601, huang2016single}, electric~\cite{Dolde2011, PhysRevApplied.16.024024,Taylor2008}, chemical~\cite{doi:10.1021/nl302979d, doi:10.1021/acsnano.9b05342, Rendler2017} and thermal signals~\cite{PhysRevLett.104.070801, chen2011temperature, PhysRevX.2.031001, nano10112282, Kucsko2013} at the nanoscale, with applications spanning condensed matter physics~\cite{casola2018probing, doi:10.1126/science.aaw4352, ku2020imaging, Bhattacharyya2024, liu2025quantum, sun2021magnetic}, quantum technology~\cite{Zu2014, gao2025signal, Childress_Hanson_2013}, and biomedicine~  \cite{doi:10.1021/acsnano.4c15003, doi:10.1021/acs.nanolett.0c04864, barnard2009diamond, schirhagl2014nitrogen, Aslam2023, yukawa2025quantum}
 . A core strength of NV centers lies in their ability to reveal vector magnetic field information through optically detected magnetic resonance (ODMR), a technique that exploits spin-dependent fluorescence to track electron spin transitions~\cite{PhysRevLett.110.130802, doi:10.1126/science.276.5321.2012}.

However, extending NV-based magnetometry to biologically relevant contexts presents unique challenges, primarily due to the random orientations of NV-containing nanodiamonds when dispersed within complex environments like cells and tissues~\cite{Kukura2009,mcguinness2011quantum, doi:10.1073/pnas.2434983100}. Previous studies have demonstrated the feasibility of orientation tracking\cite{fukushige2020identification} in biological systems using ODMR spectra, achieving sub-degree angular resolutions~\cite{mcguinness2011quantum},  full tomographic orientation reconstruction via modulated external magnetic fields~\cite{doi:10.1021/jacs.0c01191} , and simultaneous tracking of rotational and translational movement ~\cite{doi:10.1021/acs.nanolett.0c04864}
 . Despite these advances, current orientation tracking approach typically relies on the assumption that ``the external magnetic fields at their positions are well known"~\cite{SEGAWA202320}.
This interdependency---where precise magnetometry requires known orientation, and orientation determination relies on known fields---poses considerable challenges for robust orientation sensing and magnetometry when the local field signals from the target sample, such as those from ferritin~\cite{PAPAEFTHYMIOU2010886, Dubiel1999}, are present and unknown.

Our approach directly addresses this challenge. We demonstrate that a minimum of four independent ODMR measurements, under appropriately chosen bias field configurations, are sufficient to reconstruct both the full local vector magnetic field and the NV orientations. This congruence is significant: it matches the number of measurements previously established for determining only the NV axes when external magnetic fields are well-characterized~\cite{doi:10.1021/jacs.0c01191}, suggesting that comprehensive information (both local field and orientation) can be efficiently extracted with a comparable measurement overhead.

The paper is organized as follows: First, we detail the theory and methodology underpinning our approach in Secs.~\ref{sec:methodology} and \ref{sec:bias}, including the Hamiltonian for the NV center and the mathematical formulation for reconstructing orientation and field parameters from ODMR spectra under multiple bias fields. Second, we describe the experimental validations (Sec.~\ref{sec:experiment}) and numerical simulations (Sec.~\ref{sec:simulation}) of our method. This includes proof-of-concept experiments on a bulk diamond crystal with known NV orientations, followed by experiments on various single nanodiamonds where both orientation and local field are unknown. We compare experimental findings with corresponding numerical simulations for both scenarios. Finally, we conduct a noise analysis, investigating the impact of factors such as the number and strength of bias magnetic fields, and experimental noise from finite ODMR linewidth and photon shot noise. We conclude with a discussion of our findings and potential avenues for future research in Sec.~\ref{sec:conclusion}.

\begin{figure}
\centering
\includegraphics[width=1\linewidth, trim={0 1cm 0 0}, clip]{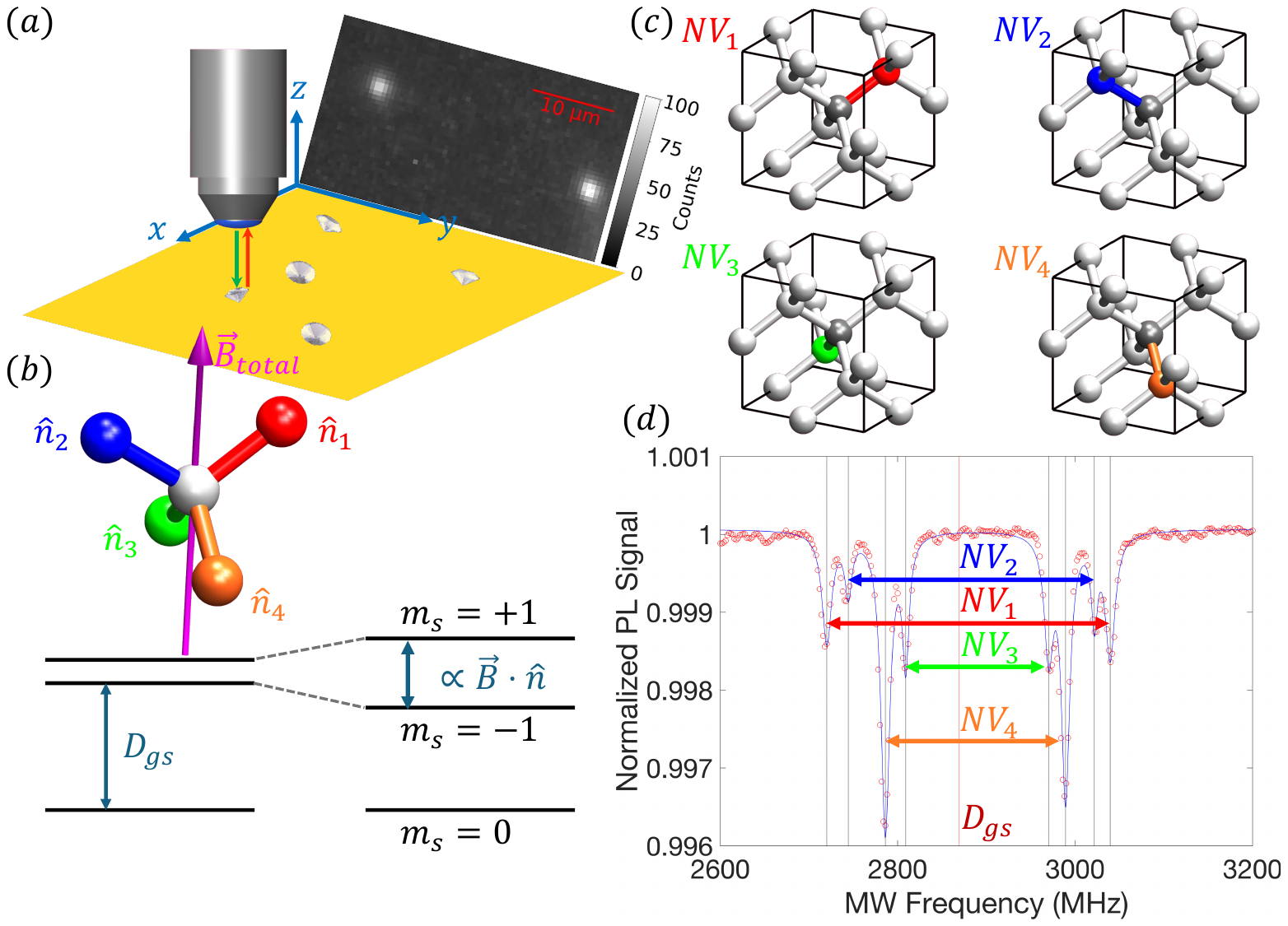}
    \caption{\textbf{Introduction to nitrogen-vacancy (NV) centers in nanodiamonds.}
   $ \mathbf{(a)}$ Schematic of an experimental setup for imaging NV centers in diamond crystals. The lab frame coordinate axes $x$, $y$, and $z$ are indicated in blue. The inset shows an optical fluorescence image of two single-crystal nanodiamond particles approximately 750~nm in diameter. When excited by a 532-nm laser, the NV centers emit red fluorescence comprising a zero-phonon line at 637~nm and a broad phonon sideband extending to approximately 800~nm. $\mathbf{(b)}$ Energy level structure of the NV center ground state under an external magnetic field. The Zeeman splitting between $m_s=\pm 1$ states is proportional (to first order) to $\vec{B}\cdot \hat{n}$, where $\hat{n}$ is the NV axis direction and $\vec{B}$ is the magnetic field at the diamond. $\mathbf{(c)}$ Atomic models of the four possible NV center orientations ($NV_1{\rightarrow}NV_4$) within the diamond lattice, each corresponding to a different crystallographic direction and colored uniquely (red, blue, green, and orange, respectively). $\mathbf{(d)}$ A typical optically detected magnetic resonance (ODMR) spectrum from a single nanodiamond showing characteristic resonance peaks corresponding to the four NV orientations, with the splitting magnitude proportional (to first order) to the projection of the magnetic field along each NV axis.}
\label{fig:intro-image}
\end{figure}

\section{Methodologies and Frameworks}\label{sec:methodology}

Having outlined the paper’s structure, we now describe the physical platform and analysis workflow in detail.
Fig~\ref{fig:intro-image} provides an overview:
  Panel (a) sketches the wide-field ODMR microscope used in this study and, in the inset, shows two representative single-crystal nanodiamond particles approximately 750~nm in diameter. Each particle hosts an ensemble of negatively charged NV centers that function as local vector magnetometers. Sample-preparation details are provided in the Supplemental Material~\cite{SupplementalMaterial}. The sensing element itself is governed by the ground-state spin-1 Hamiltonian~\cite{doherty2013nitrogen,annurev:/content/journals/10.1146/annurev-physchem-040513-103659} depicted in panel (b),
\begin{equation}
    \hat{H}=D_\mathrm{gs}\,S_z^{2} + \gamma_e\,\bigl(\vec{B}\!\cdot\!\vec{S}\bigr),
\end{equation}

where $D_\mathrm{gs} = (2\pi) \times 2.87$~GHz is the zero-field splitting, $\gamma_e = (2\pi) \times 28.0$~MHz/mT is the gyromagnetic ratio, $\vec{B}$ is the vector magnetic field, and $\vec{S}$ is the spin-1 operator for the NV center. Here we assume the magnetic field is the dominant signal and neglect the small local strain- and electric-field-induced mixing between the $|m_s=\pm1\rangle$ levels.
The resulting energy of the three NV spin levels could be obtained by solving the eigenvalues using the matrix form,
\begin{equation}
    \lambda [(D - \lambda)^2 - \mu_e^2 ( B  ^\mathrm{proj}) ^2] - (\lambda - D)\mu_e^2[(B^\mathrm{tot})^2-(B^\mathrm{proj})^2] = 0
    \label{eq:third-order}
\end{equation}
where $\lambda$ are the spin state energies, $B^\mathrm{tot}$ is the magnitude of the total magnetic field (including the applied bias magnetic field and the local field from the environment that we would like to sense), and $B^\mathrm{proj}$ is the total magnetic field's projection on the specific NV axis.

The ODMR spectrum measures the two microwave transitions from $|m_s=0\rangle$ to the $|m_s=\pm1\rangle$ levels, and the splitting between the two resonances can be calculated as $(\lambda_{+1}-\lambda_{-1})$ by fitting the spectrum using a Monte Carlo multistart fitting procedure (see Supplementary Information).
For a single-crystal nanodiamond containing an ensemble of NV centers, the NV axes can align along four crystallographic directions (Fig.~\ref{fig:intro-image}c). The resulting ODMR spectrum typically exhibits four pairs of resonances with splittings that are, to leading order, proportional to the magnetic field projection, $B^\mathrm{proj}$, onto each NV axis (Fig.~\ref{fig:intro-image}b).
We label the four NV axis as $\hat{n}_1$, $\hat{n}_2$, $\hat{n}_3$, and $\hat{n}_4$ (see supplementary equation~(S3)).

We also note that, the total magnetic field can be further decomposed into $\vec{B}^\mathrm{tot} = \vec{B}^\mathrm{bias} + \vec{B}^\mathrm{loc}$, where $\vec{B}^\mathrm{bias}$ denotes the bias magnetic field we applied externally and $\vec{B}^\mathrm{loc}$ being the local magnetic field in the environment that we want to sense. Thus the magnetic fields in the laboratory frame (labeled by coordinate \{$\hat{x}$,~$\hat{y}$,~$\hat{z}$\})
can be projected to the NV frame by:
\begin{equation}
    \hat{n}_j \cdot (\vec{B}^\mathrm{bias} + \vec{B}^\mathrm{loc}) =  {B}_j^\mathrm{proj}
    \label{eq:projection}
\end{equation}
where $j=1,2,3,4$ represents the unit vectors of the four possible NV symmetry axes.

We will apply a series of bias magnetic field and measure the corresponding ODMR splittings for all NV groups (Fig~\ref{fig:intro-image}d). By  choosing the non-coplanar bias fields like we described in Sec.~\ref{sec:bias},  plugging Eq.~\eqref{eq:projection} into Eq.~\eqref{eq:third-order} and solving these simultaneously, we can obtain the corresponding local magnetic field and nanodiamond crystallographic orientation in the lab frame.
Considering it is a set of nonlinear equations with multiple unknowns, instead of solving the equations directly, we change that into an optimization problem and opt to use MATLAB Multistart \cite{matlab_multistart}.

\begin{figure*}
      \includegraphics[width=0.9\linewidth]{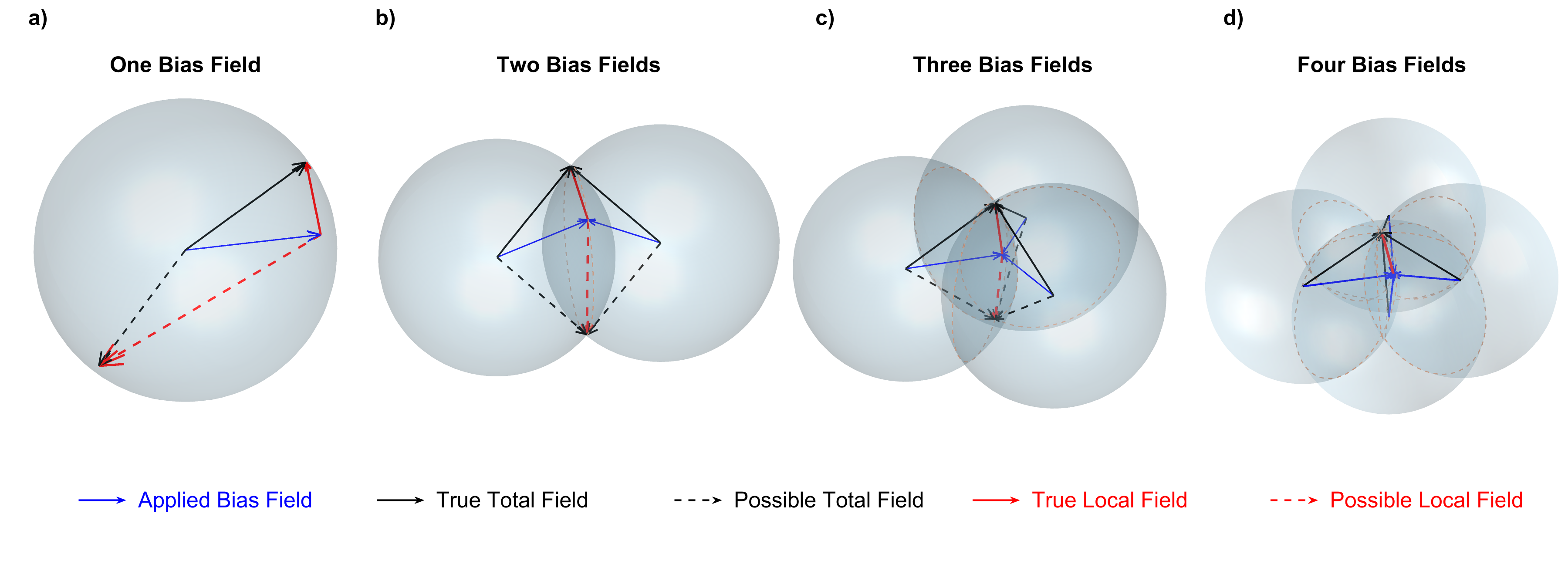}
     \caption{Illustration of the process for determining the local magnetic field and NV center orientation using multiple bias magnetic fields. \textbf{(a)} When a single bias field \( B_1^{\mathrm{bias}} \) is applied, the possible solutions of $B^\mathrm{loc}$ form a sphere, (resembled by the two red arrows), making the local field undetermined. \textbf{(b)} With a second non-parallel bias field \( B_2^{\mathrm{bias}} \), the solution space reduces to a ring (red dashed curve) at the intersection of two spheres. However, the local field is still ambiguous. \textbf{(c)} A third bias field \( B_3^{\mathrm{bias}} \) further constrains the possible solutions to two mirror-symmetric points, as shown by the solid and dashed red arrows pointing upward and downward correspondingly. This symmetry introduces a degeneracy that prevents a unique determination of the local field. \textbf{(d)} In the case of four bias fields, this degeneracy is resolved.}
    \label{fig:axes}
\end{figure*}

 \section{Minimum Number of Bias Magnetic Fields} \label{sec:bias}

A crucial aspect of our method is determining the minimum number of $\vec{B}^\mathrm{bias}$ required to extract both NV orientation and the local field, $\vec{B}^\mathrm{loc}$. Given that each NV center has a unique orientation in the nanodiamond, $\vec{B}^\mathrm{loc}$ can only be reconstructed when sufficient constraints are imposed by $\vec{B}^\mathrm{bias}$, as analyzed below.

\textbf{Single Bias Field:}
Our framework starts by setting the unknown local field $\vec{B}^\mathrm{loc}$ as a fixed vector originating at the coordinate (0,~0,~0) in the lab frame (Fig.~\ref{fig:axes}a).
When a single bias magnetic field $B^\mathrm{bias}_1$ is applied, the only directly obtainable quantity is the magnitude of the total field $|\vec{B}^\mathrm{tot}_1|$ (see Supplementary Materials), while orientation remains undetermined.
Specifically, the $\vec{B}^\mathrm{tot}_1 = \vec{B}^\mathrm{bias}_1+\vec{B}^\mathrm{loc}$ can reside anywhere on a sphere with center at (-$B^\mathrm{bias}_x$, -$B^\mathrm{bias}_y$, -$B^\mathrm{bias}_z$) and radius as the magnitude of $\vec{B}^\mathrm{tot}_1$, thus leaving the determination of $\vec{B}^\mathrm{loc}$ impossible.

\textbf{Two Bias Fields:}
By applying a second $\vec{B}^\mathrm{bias}_2$ which is not parallel to the first bias field, we impose a second constraint:
\begin{equation}
\begin{aligned}
     |\vec{B}^{\mathrm{bias}}_1 + \vec{B}^\mathrm{loc}| = |\vec{B}^{\mathrm{total}}_{1}| \\
    |\vec{B}^{\mathrm{bias}}_2 + \vec{B}^\mathrm{loc}| = |\vec{B}^{\mathrm{total}}_2|
\end{aligned}
\end{equation}
Geometrically, this restricts the set of possible $\vec{B}^\mathrm{loc}$ to a ring-shaped intersection of the two spheres, as depicted in Fig.~\ref{fig:axes}b. While this additional constraint reduces number of possible solutions, it is still not sufficient to uniquely determine $\vec{B}^\mathrm{loc}$.

\textbf{Three Bias Fields:}
Introducing a third $\vec{B}^\mathrm{bias}_3$ that is not in the same plane as the prior two bias fields, we can further restrict the possible $\vec{B}^\mathrm{loc}$ to two points, as illustrated in Fig.~\ref{fig:axes}c. This occurs because the intersection of three spheres generally results in two solutions that are symmetric about the plane formed by the centers of three spheres.

 \textbf{Four Bias Fields:}
To resolve this two-fold degeneracy, we apply a 4th non-coplanar $\vec{B}^\mathrm{bias}_4$ (i.e. the centers of the four spheres do not lie in the same plane), as shown in Fig.\ref{fig:axes}d.
Interestingly, four bias fields is also the minimum number needed to determine solely the nanodiamond axes in the absence of the unknown  $\vec{B}^\mathrm{loc}$ \cite{doi:10.1021/jacs.0c01191}.
As a result, four bias fields are sufficient to simultaneously determine $\vec{B}^\mathrm{loc}$ and nanodiamond orientation.

Subsequent numerical simulations and experimental validation confirm this necessity: with three bias fields, there remains a symmetry-induced ambiguity, while four bias fields consistently produce unique solutions for determining local magnetic fields.

\section{Experimental Validation} \label{sec:experiment}

 Two experimental configurations are used here:
\begin{itemize}
\item \textbf{A bulk diamond with known crystal orientation:} the NV axes are well-characterized, allowing us to validate the accuracy of our method in a controlled environment.
\item \textbf{Single nanodiamond:} mimicking real-world applications, where the NV axes are random and unknown. The method is tested for robustness in reconstructing these parameters.
\end{itemize}

 We first perform measurements on a [100]-cut plate diamond, with the four NV axes orientations are known \textit{a priori} and given by
\begin{equation}
    \begin{aligned}
        \hat{n}_1 & = [ +\sqrt{2},      0,  +1] / \sqrt{3}; \\
        \hat{n}_2 & = [      0,  +\sqrt{2}, -1] / \sqrt{3}; \\
        \hat{n}_3 & = [-\sqrt{2},      0,  +1] / \sqrt{3}; \\
        \hat{n}_4 & = [      0, -\sqrt{2}, -1] / \sqrt{3}.
    \end{aligned}
    \label{eq:axes}
\end{equation}
We apply   24    different bias fields , $\vec{B}^\mathrm{bias}$, each with a magnitude of approximately 1~mT using three-dimensional electromagnetic coils.
  Care was taken to ensure the applied bias fields were not coplanar to avoid degeneracies in the reconstruction.

As a calibration, we tested whether our method could accurately reconstruct these known axes and determine $\vec{B}^\mathrm{loc}$ from the measured ODMR spectra. To quantify the stability of our reconstruction from different subsets of $\vec{B}^\mathrm{bias}$, we define two metrics

(i) The standard deviation of the predicted $\vec{B}^\mathrm{loc}$:
\begin{equation}
    \delta B^\mathrm{loc}_j = \sqrt{\frac{\sum (B^\mathrm{loc}_{j} - \bar{B}^\mathrm{loc}_{j})^2}{N}}
    \label{eq:deltaB}
\end{equation}
where the summation is over a total of $N$ reconstruction runs (each using a different subset of bias fields), $j = x,~y,~z$.

(ii) The standard deviation of the predicted axes orientations represented by the great circle distance, $d_\mathrm{gc}$:
\begin{equation}
  \begin{aligned}
    d_{gc} &=  \frac{1}{N}\sum_{(\theta,\phi)} \cos^{-1}( \cos(\theta)\cos(\bar{\theta})\cos(\phi - \bar{\phi}) \\ &+ \sin(\theta)\sin(\bar\theta) );
\end{aligned}
\end{equation}
Here $(\theta, \phi)$ are the spherical coordinates representing a specific axis of NV (e.g., the first principal axis) determined in each run of reconstruction, and $(\bar{\theta}, \bar{\phi})$ are the mean spherical coordinates for that axis over the $N$ runs.

\begin{figure}
    \centering
    \includegraphics[width=\linewidth]{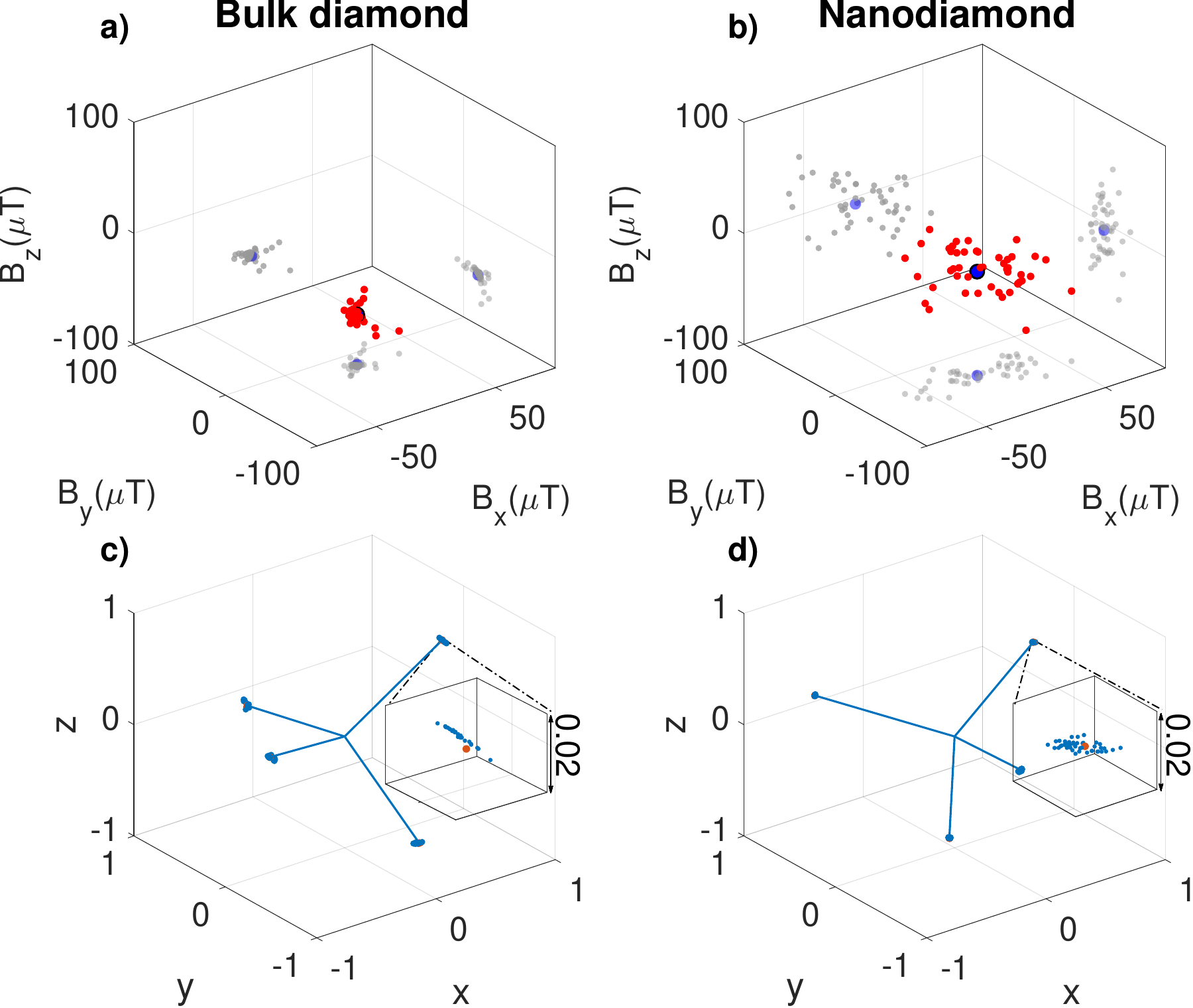}
    \caption{Reconstructed local magnetic field (a, b) and axes reconstruction (c, d) using our method. Each column represents a different scenario: bulk-diamond experiment \textbf{(a,c)} and nanodiamond experiment \textbf{(b,d)}. For each calculation shown, ten bias fields were randomly selected from the nanodiamond dataset and four from the bulk-diamond dataset. \textbf{(a,b)} Reconstructed $\vec{B}^{\mathrm{loc}}$ values (red dots) are plotted with their projections on the $yz$, $xz$, and $xy$ planes; the mean is marked by a blue dot. \textbf{(c,d)} Reconstructed NV-axis orientations (solid lines) are compared with the ground truth for the bulk diamond or with the mean orientation for the nanodiamond. The inset cubes magnify the distributions within a $0.02 \times 0.02 \times 0.02$ region around one predicted axis. Panels (a,c) show sub-$2\,\mu\mathrm{T}$ bulk-diamond field consistency and high axis fidelity, whereas panels (b,d) show approximately $50\,\mu\mathrm{T}$ nanodiamond field consistency and approximately $1^\circ$ axis consistency.}
    \label{fig:second-figure}
\end{figure}

For a proof-of-concept test, we choose $N = 50$ subsets of four randomly chosen bias field data and obtain the averaged NV axes orientations to be (Fig.~\ref{fig:second-figure})
\begin{equation}
    \begin{aligned}
        \hat{n}_1^{\mathrm{exp}} &= \hat{n}_1
        - 0.0061\,\hat{x}
        - 0.0051\,\hat{y}
        + 0.0085\,\hat{z} \\
        \hat{n}_2^{\mathrm{exp}} &= \hat{n}_2
        + 0.0013\,\hat{x}
        + 0.0095\,\hat{y}
        + 0.0137\,\hat{z} \\
        \hat{n}_3^{\mathrm{exp}} &= \hat{n}_3
        - 0.0061\,\hat{x}
        - 0.0142\,\hat{y}
        - 0.0088\,\hat{z} \\
        \hat{n}_4^{\mathrm{exp}} &= \hat{n}_4
        + 0.0108\,\hat{x}
        + 0.0098\,\hat{y}
        - 0.0135\,\hat{z}
    \end{aligned}
    \label{eq:exp_axes}
\end{equation}
The reconstructed axes agree well with the known bulk diamond crystal orientation (Eq.~\eqref{eq:axes}), with a mean angular deviation of $d_{gc}< 0.5^\circ$.

Simultaneously, we obtain the local field
$$\vec{B}^\mathrm{loc} = [-2.5, -14.9, -53.2] \pm [1.0, 1.1, 1.4]~\mu\text{T}.$$ The uncertainties represent the standard deviation ($\delta B$) across multiple reconstruction runs.
$\vec{B}^\mathrm{loc}$ closely matches the geomagnetic field in the U.S. Midwest  ~\cite{WMMHR2025}
 , highlighting the precision of our method.

 \begin{figure*}
    \centering
    \includegraphics[width=0.85\linewidth]{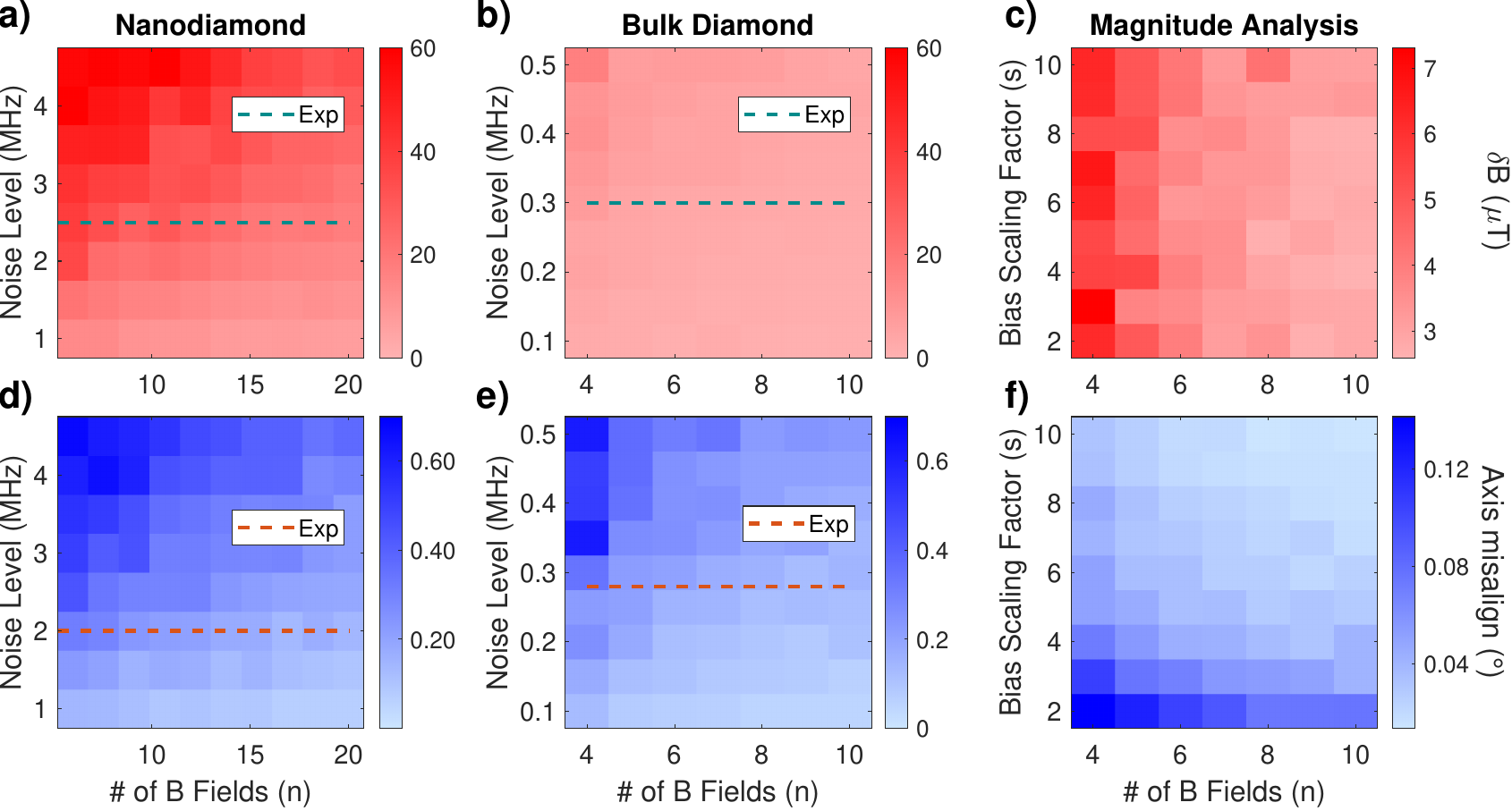}
    \caption{Impact of noise, number of bias fields, and bias-field magnitude on reconstruction. The predictions for local fields are marked in red, axes in blue, and their corresponding experimental results in green and orange. \textbf{(a,b)} Local-field deviation $\delta B$ as a function of noise level and number of bias fields for the nanodiamond and bulk-diamond simulations, respectively. \textbf{(d,e)} Axis misalignment $d_{gc}$ for the corresponding nanodiamond and bulk-diamond simulations. \textbf{(c,f)} Local-field deviation and axis misalignment, respectively, as functions of noise level and the overall bias-field scaling factor.}
    \label{fig:noise_analysis_figure}
\end{figure*}

We next extend our protocol to NV centers embedded in individual nanodiamonds, a more demanding setting where the crystal orientation is unknown \textit{a priori}.  In nanodiamonds, the ODMR resonances are substantially broadened (full-width-half-maximum, FWHM$\sim10$~MHz) compared to bulk diamond (FWHM$\sim0.6$~MHz), resulting in elevated noise floors and increased uncertainty in parameter extraction.
To mitigate this, we apply 34 different bias magnetic fields with magnitudes of approximately 5--10~mT to better separate the eight resonances. Reconstruction is then performed by randomly selecting subsets of these fields.

Using four bias fields per run, the consistency in axes prediction remains high, with $d_{gc}$ typically around 1$^\circ$. However, the prediction for the local field $\vec{B}^\mathrm{loc}$ exhibits significantly greater variation than in bulk diamond due to the broader ODMR linewidth.
To improve the field accuracy, we increase the number of bias fields per run. Fig.~\ref{fig:second-figure}b shows the average reconstructed $\vec{B}^\mathrm{loc}$ using 10 randomly selected bias field dataset, yielding $\vec{B}^\mathrm{loc} = [-7, -33, -7]\pm [36, 11, 24]\mu\mathrm{T}$ and a standard deviation $\delta B$ of approximately 50~$\mu$T.
This reduced precision arises primarily from the differing noise characteristics in bulk and nanodiamond ODMR spectra. Notably, our bulk diamond is isotopically purified to $^{13}$C concentrations below $10^{-4}$ to suppress nuclear spin noise~\cite{Balasubramanian2009}, while P1 centers and other paramagnetic defects provide additional plausible spin-bath contributions~\cite{Laraoui2013,Knowles2014,Jensen2013}. Moreover, the lower surface-to-volume ratio of bulk diamond reduces surface-spin and related defect-spin contributions relative to nanodiamonds~\cite{Tetienne2013,Lo2025}.
 As an additional cross-particle consistency check, we analyzed two different nanodiamonds on the same substrate and obtained local-field estimates that are consistent within experimental uncertainty (see Supplementary Sec.~\textit{Consistency Checks} and corresponding supplementary figures).
 \section{Numerical Analysis and Discussion} \textbf{\label{sec:simulation}}

Last, we perform simulations to quantify how experimental noise, strengths of bias fields, and number of bias fields in one run affects the reconstruction accuracy. Starting from the bulk-diamond local field of around $\vec{B}^\mathrm{loc}=50~\mu\text{T}$ pointing to -z direction, we synthetically generate eight-peak ODMR spectra and add Gaussian noise with amplitudes between 0.1 MHz and 0.5 MHz.
Each noisy spectrum is fitted to locate the resonance frequencies, which are then passed to our reconstruction algorithm.
For every noise level we repeat the procedure with 4–10 bias-field configurations randomly selected from the   24 used in the bulk-diamond experiment.
We perform an analogous analysis for nanodiamonds. Here the Gaussian noise amplitude is varied from 1 MHz to 5 MHz, and 4–20 bias-field subsets are drawn from the 34 fields available in the experiment for each run.

 The simulations uncover two clear trends (Fig.~\ref{fig:noise_analysis_figure}). First, the prediction errors, including both the local-field deviation, $\delta B$, and the angular deviation, $d_{gc}$, grow approximately linearly with the injected noise amplitude, in agreement with standard error-propagation expectations.
Second, adding more bias fields per reconstruction run improves accuracy, but the benefit saturates at around ten fields in nanodiamond case. Simulations using bulk diamond parameters exhibit the same qualitative behavior, but with uniformly smaller deviations owing to their lower noise level thus higher signal-to-noise ratio in the ODMR spectra.

 We also evaluate how bias‐field strength influences reconstruction accuracy.
Specifically, we carry out simulations in which all bias magnitudes are uniformly scaled by factors of 2–10 (denoted “Bias-Scaling Factor” in panels (c) and (f) of Fig.~\ref{fig:noise_analysis_figure}). Increasing the field strength noticeably enhances axis determination, yet yields only marginal improvement in the local-field error.
This behavior can be intuitively understood using the geometrical picture in Fig.~\ref{fig:axes}. Axis orientation is inferred indirectly from the relative directions of $\vec{B}^\mathrm{bias}$; with a fixed noise floor, scaling the bias magnitudes reduces the angular uncertainty and thus yields a more precise axis estimate. By contrast, the local field $\vec{B}^\mathrm{loc}$ is extracted from the intersection point of the total field vectors. Once the bias field is already much larger than $\vec{B}^\mathrm{loc}$, further scaling hardly changes the absolute positional error as it is set by the same noise floor.

\section{Conclusions \& Outlooks}\label{sec:conclusion}

We have developed and experimentally validated a framework for simultaneously determining the NV center orientation and the local vector magnetic field.
Our results show that at least four non-coplanar bias fields are required, echoing with previous orientation-only methods but uniquely enable vector field reconstruction.
The primary performance bottleneck of our protocol arises from the intrinsic noise inside nanodiamond sensors: $^{13}$C nuclear spin bath, strain variations, and structural inhomogeneities. Addressing these noise sources is critical to improving the accuracy and reliability of our method under real-world conditions; approaches outlined in Refs.~\cite{PhysRevApplied.10.034044,PhysRevApplied.14.014055,PhysRevApplied.22.044030,dubey2025magneticfieldorientationdependence,Yamamoto_2025,PhysRevB.84.195204, PhysRevA.96.042115, doi:10.1126/science.1192739} offer viable solutions.
In this experiment, we focus on the quasi-static regime, in which the nanodiamond orientation and local field are approximately constant over the spectra used for one reconstruction. Although the local field may depend on the applied bias field, even for ferritin-like moments a low-field estimate suggests that bias-induced alignment is small under the present room-temperature, millitesla-field conditions~\cite{Gider1995,PAPAEFTHYMIOU2010886}; systematically quantifying this dependence for other magnetic species and operating conditions is left for future work.
Ultimately, we anticipate a robust, simultaneous readout of nanodiamond orientation and local magnetic fields will enable new classes of experiments, from tracking ferritin-related magnetic signatures in living systems \cite{PAPAEFTHYMIOU2010886, Dubiel1999} to diagnosing nanoscale magnetism in quantum materials.

\begin{acknowledgments}
The authors acknowledge the support of NSF Grant 2503230 for enabling this research. C.Zu acknowledges support from NSF Expand-QISE 2328837. C.Zhang acknowledges support from NSF ECCS-2411394.

\end{acknowledgments}

\bibliography{apssamp}

\end{document}


\preprint{APS/123-QED}

\title{Supplementary Material for Simultaneous Determination of Local Magnetic Fields and Sensor Orientation with Nitrogen-Vacancy Centers in Nanodiamond}

\author{Yizhou Wang}
\thanks{These authors contributed equally to this work.}
\author{Haochen Shen}
\thanks{These authors contributed equally to this work.}
\author{Zhongyuan Liu}
\author{Yue Yu}
\affiliation{Department of Physics, Washington University in St. Louis, St. Louis, MO, USA}

\author{Shengwang Du}

\affiliation{Elmore Family School of Electrical and Computer Engineering,
Purdue University, West Lafayette, IN, USA}
\affiliation{Department of Physics and Astronomy, Purdue University, West Lafayette, IN, USA}
\affiliation{Purdue Quantum Science and Engineering Institute,
Purdue University, West Lafayette, IN, USA}
\author{Chong Zu}
\email{zu@wustl.edu}
\affiliation{Department of Physics, Washington University in St. Louis, St. Louis, MO, USA}
\affiliation{Center for Quantum Leaps, Washington University, St. Louis, MO, USA}

\author{Chuanwei Zhang}
\email{chuanwei.zhang@wustl.edu}
\affiliation{Department of Physics, Washington University in St. Louis, St. Louis, MO, USA}
\affiliation{Center for Quantum Leaps, Washington University, St. Louis, MO, USA}

\maketitle

\section{Derivation of NV Center Eigenvalue Equation}
\label{sec:appendix_eigenvalue_derivation}

The ground state Hamiltonian for an NV center in an external magnetic field $\vec{B}$ is given by Eq.~(1) in the main text:

\begin{equation}
   \hat{H}=D_\mathrm{gs}\,S_z^{2} + \gamma_e\,\bigl(\vec{B}\!\cdot\!\vec{S}\bigr),
\end{equation}

In the $m_s = \{|+1\rangle, |0\rangle, |-1\rangle\}$ basis, the spin matrices are ($\hbar=1$):
\begin{widetext}
\[
S_x = \frac{1}{\sqrt{2}} \begin{pmatrix} 0 & 1 & 0 \\ 1 & 0 & 1 \\ 0 & 1 & 0 \end{pmatrix}, \quad
S_y = \frac{1}{i\sqrt{2}} \begin{pmatrix} 0 & 1 & 0 \\ -1 & 0 & 1 \\ 0 & -1 & 0 \end{pmatrix}, \quad
S_z = \begin{pmatrix} 1 & 0 & 0 \\ 0 & 0 & 0 \\ 0 & 0 & -1 \end{pmatrix}
\]
\end{widetext}
Assuming the magnetic field is $\vec{B} = (B_x, B_y, B_z)$, the Hamiltonian matrix becomes (setting $h=1$):
\begin{widetext}

\begin{equation}
    H = \begin{pmatrix}
        D + \gamma_{e} B_z & \frac{\gamma_{e}}{\sqrt{2}}(B_x - iB_y) & 0 \\
        \frac{\gamma_{e}}{\sqrt{2}}(B_x + iB_y) & 0 & \frac{\gamma_{e}}{\sqrt{2}}(B_x - iB_y) \\
        0 & \frac{\gamma_{e}}{\sqrt{2}}(B_x + iB_y) & D - \gamma_{e} B_z
\end{pmatrix}
\end{equation}

\label{eq:Hamiltonian}
\end{widetext}

Let $\omega_z = \gamma_{e} B_z$ and $\omega_\perp = \gamma_{e} \sqrt{B_x^2 + B_y^2}$. The characteristic equation $\det(H - \lambda I) = 0$ yields the cubic form to the form used in Eq.~(1) of the main text:
\[
\lambda [(D - \lambda)^2 - \gamma_{e}^2(B_{j}^{\mathrm{proj}})^2] - (\lambda - D)\gamma_{e}^2[({B^{\mathrm{total}}})^{2}-(B_{j}^{\mathrm{proj}})^2] = 0
\]
Solving this equation provides the energy eigenvalues $\lambda_0, \lambda_{+1}, \lambda_{-1}$, where in the main text, we will take advantage of the differences between $\lambda_{+1}, \lambda_{-1}$. given that they would be less susceptible to the shift of center frequencies.

\subsection{Derivation of NV Axes Parameterization}
\label{sec:appendix_axes_derivation}

Equation \eqref{eq:axes-exp} parameterizes the four NV axes $\{\hat{n}_{j}\}$ using the spherical coordinates $(\theta_1, \phi_1)$ of $\hat{n}_1$ and an auxiliary angle $\alpha$, satisfying the tetrahedral constraint $\hat{n}_{k} \cdot \hat{n}_l = -1/3$ ($k \neq l$).

The derivation involves:
\begin{enumerate}
    \item Defining $\hat{n}_1 = (\cos\theta_1\cos\phi_1, \cos\theta_1\sin\phi_1, \sin\theta_1)$.
    \item Defining an orthonormal basis $(\hat{u}, \hat{v}, \hat{n}_1)$ relative to $\hat{n}_1$:
    \begin{align*}
        \hat{u} &= (-\sin\phi_1, \cos\phi_1, 0) \\
        \hat{v} &= (-\sin\theta_1\cos\phi_1, -\sin\theta_1\sin\phi_1, \cos\theta_1)
    \end{align*}
    \item Expressing $\hat{n}_{i}$ ($i=2,3,4$) as $\hat{n}_{j} = c_1 \hat{n}_1 + c_{u,j} \hat{u} + c_{v,j} \hat{v}$.
    \item Applying constraints: $\hat{n}_{j} \cdot \hat{n}_1 = -1/3 \implies c_1 = -1/3$. And $\hat{n}_{j} \cdot \hat{n}_{j} = 1 \implies c_{u,j}^2 + c_{v,j}^2 = 8/9$. Also $\hat{n}_{k} \cdot \hat{n}_l = -1/3 \implies c_{uk}c_{ul} + c_{vk}c_{vl} = -4/9$ for $i \neq j$.
    \item Parameterizing the $(c_{u,j}, c_{v,j})$ components using the angle $\alpha$ and $2\pi/3$ rotations to satisfy the above conditions:
    \begin{align*}
        c_{u,j} &= \frac{2\sqrt{2}}{3} \sin\left(\alpha + \frac{2\pi}{3}(j-2)\right) \\
        c_{v,j} &= \frac{2\sqrt{2}}{3} \cos\left(\alpha + \frac{2\pi}{3}(j-2)\right)
    \end{align*}
    \item Substituting back into $\hat{n}_{j} = c_1 \hat{n}_1 + c_{u,j} \hat{u} + c_{v,j} \hat{v}$ yields the expression in Eq.\eqref{eq:axes-exp}.
\end{enumerate}

\begin{widetext}
    \begin{equation}
    \hat{n}_{j} =
        \begin{cases}
            (\cos\theta_1\cos\phi_1,\, \cos\theta_1\sin\phi_1,\, \sin\theta_1), & j=1, \\[1mm]\\
            -\dfrac{1}{3}(\cos\theta_1\cos\phi_1,\, \cos\theta_1\sin\phi_1,\, \sin\theta_1) \\ \quad
            -\dfrac{2\sqrt{2}}{3}\,\sin\!\Bigl(\alpha+\dfrac{2\pi}{3}(j-2)\Bigr)(-\sin\phi_1,\, -\cos\phi_1,\, 0) \\[1mm]
            \quad -\dfrac{2\sqrt{2}}{3}\,\cos\!\Bigl(\alpha+\dfrac{2\pi}{3}(j-2)\Bigr)(-\sin\theta_1\cos\phi_1,\, -\sin\theta_1\sin\phi_1,\, \cos\theta_1), & j=2,3,4,
        \end{cases}
        \label{eq:axes-exp}
    \end{equation}

\end{widetext}

\subsection{Optimization Problem Formulation}
\label{sec:appendix_optimization}

Determining the six parameters $(\theta_1, \phi_1, \alpha, B_x^{\mathrm{loc}}, B_y^{\mathrm{loc}}, B_z^{\mathrm{loc}})$ from noisy experimental ODMR splittings is formulated as a least-squares minimization problem, suitable for solvers like MATLAB Multistart. The objective is to minimize the cost function:
\begin{widetext}
\begin{equation}
    \text{Cost}(\theta_1, \phi_1, \alpha, \vec{B}^{\mathrm{loc}}) = \sum_{j=1}^{N_{\text{fields}}} \sum_{j=1}^{4}  w_{ij} \left[ S_{ij}^{\text{measured}} - S_{ij}^{\text{calculated}}(\theta_1, \phi_1, \alpha, \vec{B}^{\mathrm{loc}}, \vec{B}^{\mathrm{bias}}_i) \right]^2
    \label{eq:cost_function_appendix_revised}
\end{equation}
\end{widetext}
Here, $i$ indexes the $N_{\text{fields}}$ bias fields $\vec{B}^{\mathrm{bias}}_i$, $j$ indexes the 4 NV axes $\hat{n}_j$. $S_{ij}^{\text{measured}}$ are the splittings measured in experiments, and $S_{ij}^{\text{calculated}}$ are derived from the trial parameters by:

\begin{enumerate}
    \item \textbf{Total Magnetic Field Calculation:} For each applied bias field configuration $i$, evaluate the total magnetic field as
\begin{equation}
    \vec{B}^{\mathrm{total}}_i = \vec{B}^{\mathrm{bias}}_i + \vec{B}^{\mathrm{loc}},
\end{equation}
where $\vec{B}^{\mathrm{loc}}$ is the unknown local field.

\item \textbf{Field Projection and Magnitude:} Compute both the projection of the total field along the axis $\hat{n}_j$ using Eq.~\eqref{eq:axes-exp} and its magnitude:
\begin{align}
    B_{ji}^{\mathrm{proj}} &= \vec{B}^{\mathrm{total}}_i \cdot \hat{n}_j, \\
    |\vec{B}^{\mathrm{total}}_i| &= \left\| \vec{B}^{\mathrm{total}}_i \right\|.
\end{align}

\item \textbf{Energy Level Calculation:} Solve the eigenvalue equation (Eq.~(2))) to obtain the three spin-1 eigenenergies: $\lambda_0$, $\lambda_{\pm 1}$.

\item \textbf{Transition Splitting:} Determine the expected splitting between the $m_s = \pm1$ states:
\begin{equation}
    S_{ij}^{\text{calculated}} = |\lambda_1 - \lambda_{-1}|.
\end{equation}

\end{enumerate}

In our case $w_{ij}$ is set to be {$10^{14}$} such that the optimization in MATLAB could proceed robustly given that we used GHz as the unit of frequencies. The optimization searches for the parameters minimizing this cost, subject to angular constraints ($\theta_1 \in [0, \pi]$, $\phi_1, \alpha \in [0, 2\pi]$) and potential bounds on $\vec{B}^{\mathrm{loc}}$, given that the optimization always needs a upper and lower bound. This is also where three bias fields could also potentially determine the local magnetic field, given that in a specific setup for bias fields, the "spurious result" could be say 2 mT away from the true result that we're looking for, thus determining the real result we're looking for.

\subsection{Determination of Total Magnetic Field Magnitude}
\label{sec:appendix_total_field}

A key aspect of the analysis pipeline involves determining the magnitude of the total magnetic field, $|\vec{B}^{\mathrm{total}}_i| = |\vec{B}^{\mathrm{bias}}_i + \vec{B}_l|$, for each applied bias field $\vec{B}^{\mathrm{bias}}_i$. This magnitude is needed, for instance, in the eigenvalue calculation Eq.~(5). Crucially, $|\vec{B}^{\mathrm{total}}_i|$ can be determined directly from the ODMR splittings corresponding to the four NV axes, without prior knowledge of the specific NV orientation (i.e., the angles $\theta_1, \phi_1, \alpha$). This justifies determining the total field magnitudes as an initial step.

The relationship between the total field $\vec{B}^{\mathrm{total}}$ and its projections $B^{\mathrm{proj}}_{j} = \vec{B}^{\mathrm{total}} \cdot \hat{n}_{j}$ along the four NV axes $\hat{n}_{j}$ allows this determination. The projected field magnitudes $B^{\mathrm{proj}}_{ij}$ are directly related to the measured ODMR splittings via the eigenvalue equation (Eq.~(5) or its equivalent solution).

Consider the sum of the squares of these projections:
\[
\sum_{j=1}^4 (B^{\mathrm{proj}}_{ij})^2 = \sum_{j=1}^4 (\vec{B}_{i}^{\mathrm{total}} \cdot \hat{n}_j)^2
\]
Let the total magnetic field vector be $\vec{B}^{\mathrm{total}} = (B_x^{\mathrm{total}}, B_y^{\mathrm{total}}, B_z^{\mathrm{total}})$. Expanding the dot product squared gives:
\[
\sum_{j=1}^4 (B^{\mathrm{proj}})_{j}^2 = \sum_{j=1}^4 \left( B_x^{\mathrm{total}} (\hat{n}_{j})_x + B_y^{\mathrm{total}} (\hat{n}_{j})_y + B_z^{\mathrm{total}} (\hat{n}_{j})_z \right)^2
\]
Expanding the square and rearranging the summation yields terms involving sums over the axis components:
\begin{widetext}
    \begin{align*}
        \sum_{j=1}^4 (B^{\mathrm{proj}}_{j})^2 = &(B_x^{\mathrm{total}})^2 \sum_{j=1}^4 (\hat{n}_{j})_x^2 + (B_y^{\mathrm{total}})^2 \sum_{j=1}^4 (\hat{n}_{j})_y^2 + (B_z^{\mathrm{total}})^2 \sum_{j=1}^4 (\hat{n}_{j})_z^2 \\
        &+ 2 B_x^{\mathrm{total}} B_y^{\mathrm{total}} \sum_{j=1}^4 (\hat{n}_{j})_x (\hat{n}_{j})_y + 2 B_x^{\mathrm{total}} B_z^{\mathrm{total}} \sum_{j=1}^4 (\hat{n}_{j})_x (\hat{n}_{j})_z + 2 B_y^{\mathrm{total}} B_z^{\mathrm{total}} \sum_{j=1}^4 (\hat{n}_{j})_y (\hat{n}_{j})_z
    \end{align*}
\end{widetext}

Due to the tetrahedral symmetry of the four NV axes $\{\hat{n}_1, \hat{n}_2, \hat{n}_3, \hat{n}_4\}$, regardless of their specific orientation in the laboratory frame, the following identities hold:
\begin{gather}
    \sum_{j=1}^4 (\hat{n}_{j})_k^2 = \frac{4}{3} \quad \text{for } k = x, y, z \label{eq:sum_sq_comp} \\
    \sum_{j=1}^4 (\hat{n}_{j})_k (\hat{n}_{j})_l = 0 \quad \text{for } k \neq l \label{eq:sum_cross_comp}
\end{gather}
Eq~\eqref{eq:sum_sq_comp} for $k=z$ was explicitly shown to yield $4/3$ independent of $\theta_1$ and $\alpha$ in the calculation leading to $S=4/3$. By symmetry, the same result holds for the $x$ and $y$ components. The vanishing of the cross terms in Eq.~\eqref{eq:sum_cross_comp} also arises from this symmetry.

Substituting these identities (Eqs.~\eqref{eq:sum_sq_comp} and \eqref{eq:sum_cross_comp}) into the expansion for $\sum (B^{\mathrm{proj}}_{j})^2$:
\[
\sum_{j=1}^4 (B^{\mathrm{proj}}_{j})^2 = (B_x^{\mathrm{total}})^2 \left(\frac{4}{3}\right) + (B_y^{\mathrm{total}})^2 \left(\frac{4}{3}\right) + (B_z^{\mathrm{total}})^2 \left(\frac{4}{3}\right) + 0 + 0 + 0
\]
\[
\sum_{j=1}^4 (B^{\mathrm{proj}}_{j})^2 = \frac{4}{3} \left[ (B_x^{\mathrm{total}})^2 + (B_y^{\mathrm{total}})^2 + (B_z^{\mathrm{total}})^2 \right] = \frac{4}{3} |\vec{B}^{\mathrm{total}}|^2
\]
Rearranging this gives the relationship as:
\begin{equation}
    |\vec{B}^{\mathrm{total}}|^2 = \frac{3}{4} \sum_{j=1}^4 (B^{\mathrm{proj}}_{j})^2 = \frac{3}{4} \left[ (B_{1}^{\mathrm{proj}})^2 + (B_{2}^{\mathrm{proj}})^2 + (B_{3}^{\mathrm{proj}})^2 + (B_{4}^{\mathrm{proj}})^2 \right]
    \label{eq:A3_derived}
\end{equation}
Since the derivation relies only on the tetrahedral symmetry reflected in Eqs.~\eqref{eq:sum_sq_comp} and \eqref{eq:sum_cross_comp}, which are independent of the specific orientation angles $(\theta_1, \phi_1, \alpha)$, Eq.~\eqref{eq:A3_derived} holds universally.

Therefore, by determining the four projected field magnitudes $B^{\mathrm{proj}}_{i}$ from the measured ODMR splittings for a given applied bias field $\vec{B}^{\mathrm{bias}}_i$, one can calculate the squared magnitude of the total magnetic field $|\vec{B}^{\mathrm{total}}_i|^2$ using Eq.~\eqref{eq:A3_derived} \textit{without} needing to know the axes' orientation first. As a thought experiment, we can start with a good initial guess of Zeeman splitting first, then obtain the total magnetic field, and then calculate the $B^{\mathrm{proj}}_i$ and $B^{\mathrm{total}}$ recursively to obtain the correct result. This justifies the procedure where total field magnitudes can be determined prior to solving the full optimization problem for the axes and the local field vector.

\section{Additional Information on Geometric Interpretation of Field Determination and Degeneracy}
\label{sec:appendix_geometry}

The process of determining the unknown local magnetic field $\vec{B}^{\mathrm{loc}}$ relies on constraining its possible values by applying known bias fields $\vec{B}^{\mathrm{bias}}_i$ and measuring the resulting total field magnitudes $|\vec{B}^{\mathrm{total}}_i|$. Each measurement imposes a constraint:
\begin{equation}
    |\vec{B}^{\mathrm{bias}}_i + \vec{B}^{\mathrm{loc}}| = |\vec{B}^{\mathrm{total}}_i|
    \label{eq:constraint_sphere_revised}
\end{equation}

Equation~\eqref{eq:constraint_sphere_revised} has a simple geometric interpretation. For a given applied bias field $\vec{B}^{\mathrm{bias}}_i$, the unknown local field $\vec{B}^{\mathrm{loc}}$ must lie on a sphere centered at $-\vec{B}^{\mathrm{bias}}_i$. The radius of this sphere is the measured total-field magnitude $|\vec{B}^{\mathrm{total}}_i|$. Repeating the measurement with different bias fields produces several such spheres. Their common intersection gives the possible values of $\vec{B}^{\mathrm{loc}}$ and, with sufficient independent bias fields, identifies the true local field.

The progression of determining $\vec{B}_l$ and the emergence of degeneracies is illustrated by the schematics in Fig.~2 and by numerical simulations of the possible solutions below.

\begin{figure}
    \includegraphics[width=0.6\textwidth]{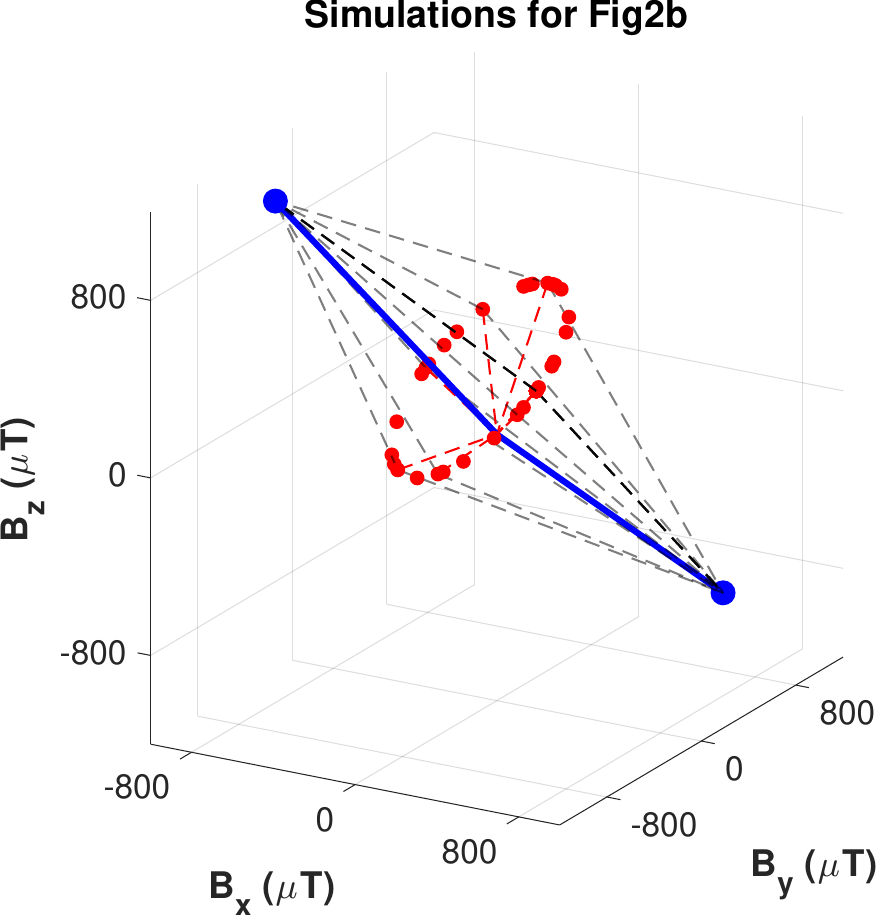}
    \includegraphics[width=0.6\textwidth]{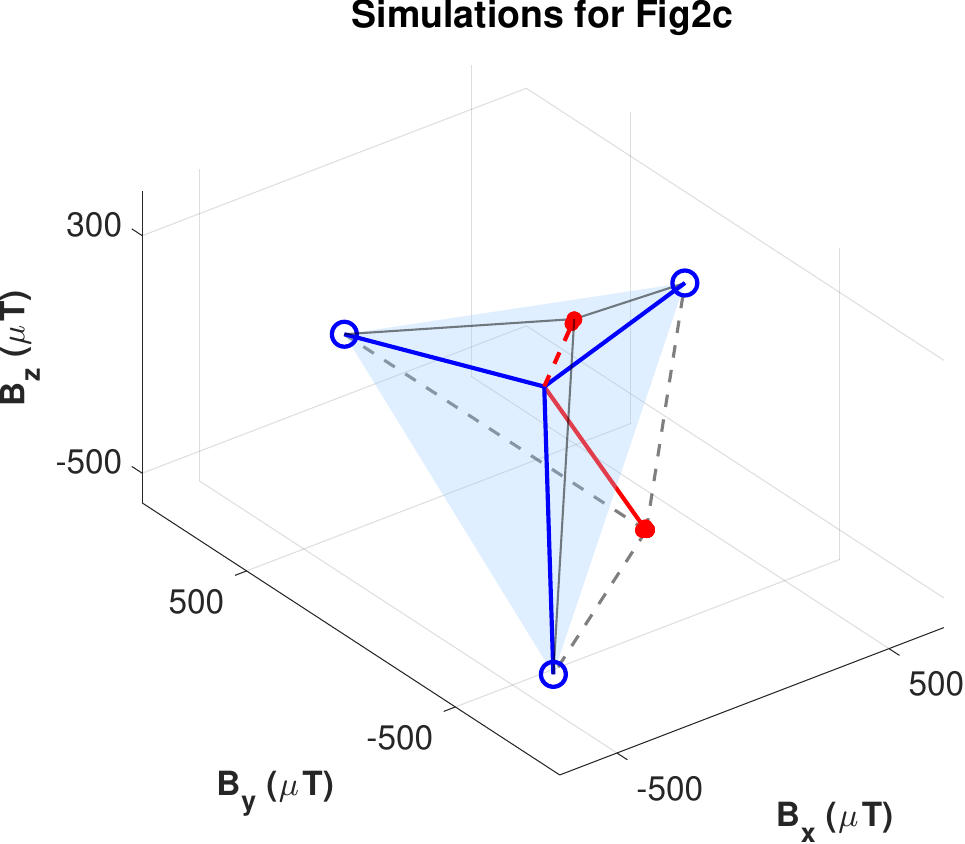}
    \caption{Simulations for local magnetic fields predictions with two and three bias magnetic fields. The blue lines and the blue dots represent the bias magnetic fields, the red dots together with the dashed red lines represent the predicted local magnetic fields, and the dashed gray lines representing the total magnetic fields. \textbf{a)} With two bias magnetic fields, the local magnetic is determined on a ring, and the two blue points (i.e. the total magnetic fields) are mirrorly symmetric across the plane formed by the predicted local magnetic fields. \textbf{b)} A light blue plane denote the plane formed by the three bias magnetic fields, and the local magnetic fields still has a double degeneracy. Those two points are mirrorly symmetric across the light blue planes, spanned by the three blue dots.}
    \label{fig:sup_w_mag}
\end{figure}

\begin{figure}
    \centering
    \includegraphics[width=0.7\linewidth]{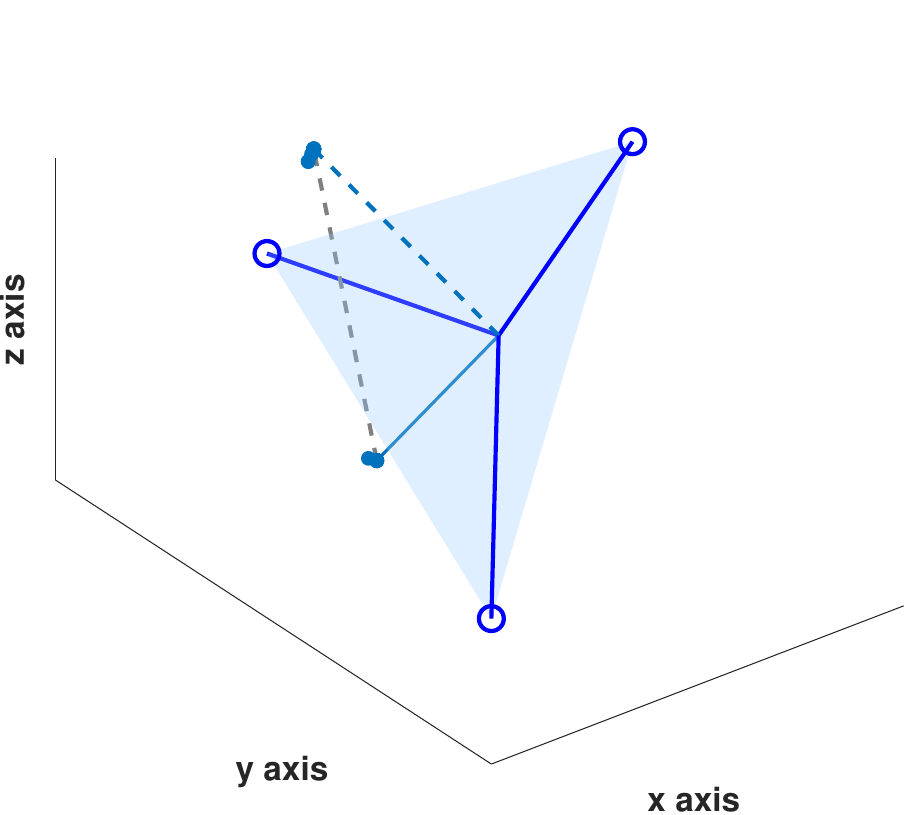}
    \caption{Axis predictions for three bias fields. The reference axes are shown as solid blue lines with open circles, and the alternative predicted axes are shown as dashed lines with filled points. The two sets are mirror-symmetric across the light-blue plane.}
    \label{fig:axes_Three_Fields}
\end{figure}

\begin{itemize}
    \item \textbf{One Bias Field ($N=1$):} With a single measurement ($i=1$), Eq.~\eqref{eq:constraint_sphere_revised} defines one sphere $S_1$. The endpoint of the local magnetic field vector $\vec{B}_l$ can lie anywhere on this surface (Fig.~2a). The system is severely underdetermined. As for the axes, a rough understanding could be obtained through Eq~(5), where the splittings are only sensitive to the total magnetic fields and the magnetic field parallel to the bias fields. So for a settled total field, the splitting could be determined using only the magnetic field parallel to the axes, i.e. $\hat{n}_{j}\cdot \vec{B}^{\mathrm{total}}$, which gives all those four axes a rotational freedom (Imagine a case where the total field points to z direction in the equation \eqref{eq:axes-exp}. We can clearly see that only $\alpha$ and $\theta$ could be determined in such case, leaving a common rotational freedom for all four axes). We could also see that the properties of axes could only be expressed by the direction of the total vector magnetic field, while the properties of the magnetic field (its magnitude) is not dependent on the axes at all, so we can say that only the magnitude of the total field is determined.

    \item \textbf{Two Bias Fields ($N=2$):} Applying a second, non-parallel bias field $\vec{B}^{\mathrm{bias}}_2$ yields a second sphere $S_2$. Possible solutions for $\vec{B}_l$ lie on the intersection $S_1 \cap S_2$, which forms a ring (Fig.~2b). The solution thus remains ambiguous. To numerically demonstrate this, a simultaneous determination of local magnetic fields and axes was conducted with two bias magnetic fields. Figure \ref{fig:sup_w_mag}a presents a case where only two bias fields are utilized to calculate the local magnetic fields, clearly showing a ring of possible values for the local magnetic field, as suggested in the main text.

    Regarding the determination of the axes, although not explicitly figured for the two-bias-fields case, the existence of a set of different possible local magnetic fields implies that the corresponding axes are also not uniquely determined. This can be understood as follows: for given total magnetic field measurements and known bias fields, a set of congruent triangles is formed. Consequently, the angles involving the local magnetic field are the same for any predicted $\vec{B}_l$ on the ring. Therefore, if a solved local magnetic field is rotated by a certain angle from an initial position, its corresponding axes can also be rotated by the same angle to yield identical solutions.

    \item \textbf{Three Bias Fields ($N=3$):} A third bias field $\vec{B}^{\mathrm{bias}}_3$, whose sphere center is not collinear with the centers of the first two spheres, introduces a third sphere $S_3$. The three sphere centers define a plane, and the intersection $S_1 \cap S_2 \cap S_3$ generally yields two distinct points related by reflection across that plane (Fig.~2c). In practice, this degeneracy can manifest as shown in Fig.~2b, where the two possible local magnetic fields are located as mirror images of each other across the plane formed by the centers of the three spheres. It is still possible to determine the local magnetic field in such a setup, provided there is prior knowledge of its approximate magnitude. In such a scenario, the bias magnetic fields can be configured such that the spurious result yields a local magnetic field prediction significantly different (e.g., at least an order of magnitude larger) from the true local field.

    For the three-bias-fields case, the determination of the axes is further illustrated in Fig.~\ref{fig:axes_Three_Fields}. The predicted axes can be either the true axes or the spurious set, which are mirror-symmetric with respect to the shaded blue plane, analogous to the local magnetic field solutions.

    \item \textbf{Four Bias Fields ($N=4$):} A fourth bias field $\vec{B}^{\mathrm{bias}}_4$ is required to obtain a unique solution. The center of the sphere corresponding to this field, $-\vec{B}^{\mathrm{bias}}_4$, must be non-coplanar with the centers of the first three spheres $-\vec{B}^{\mathrm{bias}}_1, -\vec{B}^{\mathrm{bias}}_2, -\vec{B}^{\mathrm{bias}}_3$. This fourth sphere, $S_4$, intersects only one of the two points derived from the three-field case, thereby breaking the symmetry and yielding a unique solution for $\vec{B}_l$.
\end{itemize}

This geometric analysis provides supplementary justification for requiring four non-coplanar bias fields.

\section{Determination and choice of bias fields}

Given that sometimes the coils aren't calibrated, the bias fields also cannot be determined. In that case the following approach could be used to help determine the bias fields.
\begin{itemize}
    \item Firstly take a set of data. On top of our original code, we also set the linear bias fields-current relations with different coil as unknown parameters
    \item Use all these data (Or some of them, depending on the ODMR profile), to calculate the fields-current relations, axes, and local magnetic fields concurrently.
\end{itemize}

 In practice, we calibrate using a large subset of spectra chosen to maintain stable global fitting while preserving broad directional coverage of the applied bias configurations.  We apply our bias fields with three sets of coils and adjust the current in each coil such that   the  bias magnetic fields are usually chosen to be much larger than the noise levels, for example $\sim$ 100 times.  We also prefer larger bias-field magnitudes, since they improve spectral separability and thus increase reconstruction consistency for both local-field and orientation estimates.  Also, it is suggested that the four bias fields should be far from each other, for example $\hat{n}^{bias}_{B_i}\cdot \hat{n}^{bias}_{B_j} \approx -1/3$, so as to get  much better predictions in terms of local magnetic fields.

 For the bulk-diamond measurements reported in this work, we use the coil currents $(I_x,I_y,I_z)$ as the control parameters and describe the bias field by an approximately linear relation
\begin{equation}
    \vec{B}^{\mathrm{bias}} \approx I_x \vec{B}_x + I_y \vec{B}_y + I_z \vec{B}_z,
\end{equation}
where the fitted coil-basis vectors (in Tesla per unit current) are
\begin{equation}
    \begin{aligned}
        \vec{B}_x &= [0.46,\,-0.00,\,0.00] \times 10^{-3},\\
        \vec{B}_y &= [0.00,\,0.556,\,-0.01] \times 10^{-3},\\
        \vec{B}_z &= [0.00,\,0.00,\,-0.485] \times 10^{-3}.
    \end{aligned}
\end{equation}

 For the nanodiamond measurements, we similarly use a fitted linear relation between applied currents and bias fields. The fitted coil-basis vectors (in Tesla per unit current) are
\begin{equation}
    \begin{aligned}
    \vec{B}_x &= [1.22,\,0.00,\,0.00] \times 10^{-3}, \\
    \vec{B}_y &= [-1.01,\,5.74,\,0.00] \times 10^{-3}, \\
    \vec{B}_z &= [-0.42,\,-1.46,\,3.47] \times 10^{-3}.
    \end{aligned}
\end{equation}

 \subsection{Impact of number of peaks on the fitting results}
The mechanism that broadening could impact the results of our fitting is intuitive, given that peak widths are usually the reason why we have error bars in the spectra. However, it is also very reasonable to ask another question: Given that we actually know the physics, could we say determine the peak positions beyond the measuring accuracies because we can infer the positions of the peaks by fitting? This is a reasonable argument, and in reality, when we measure those spectra with eight peaks in total, we can indeed get much better results than what we presented in our main manuscript, as seen in \ref{fig:demo-eight-peaks}:

\begin{figure}
    \centering
    \includegraphics[width=\linewidth]{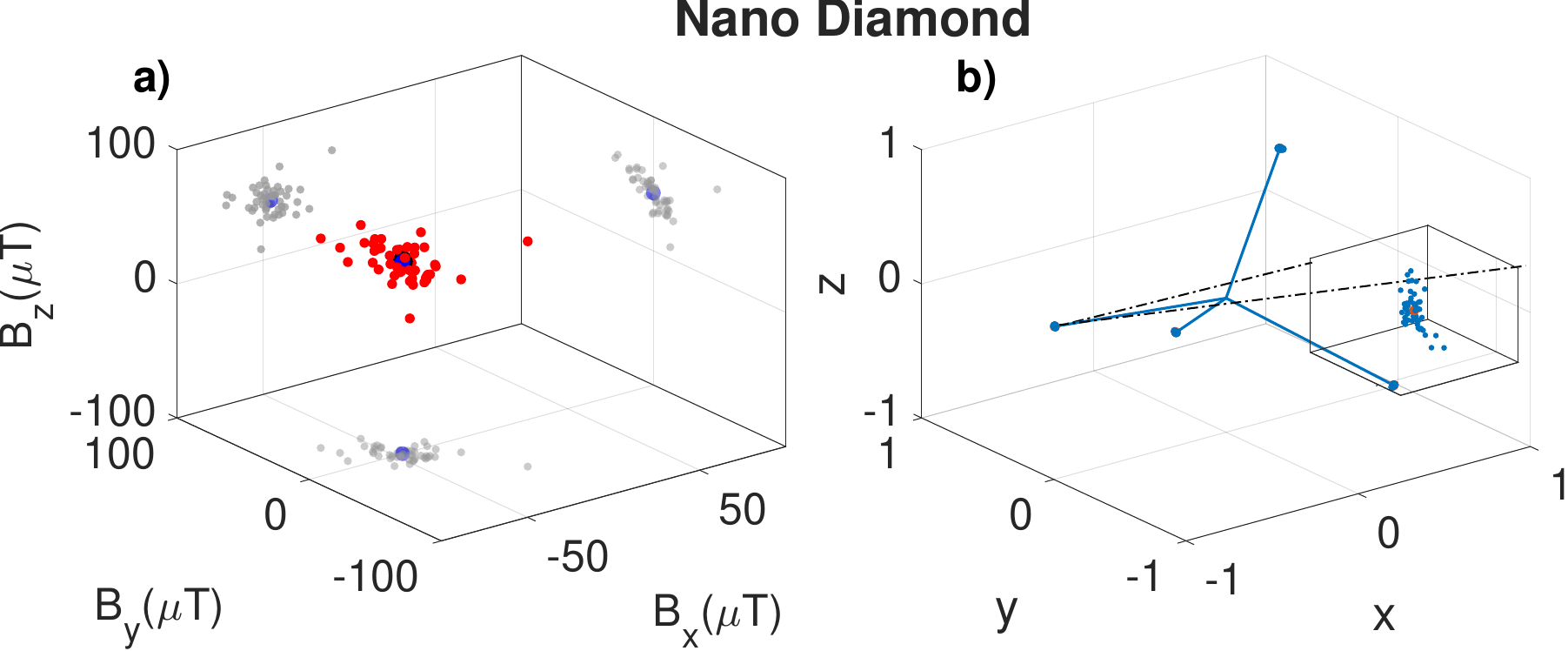}
    \caption{Local magnetic fields and NV axes fitted using four bias-field configurations and ODMR spectra with eight resolved peaks.}
    \label{fig:demo-eight-peaks}
\end{figure}

\begin{figure}
    \centering
    \includegraphics[width=0.8\linewidth]{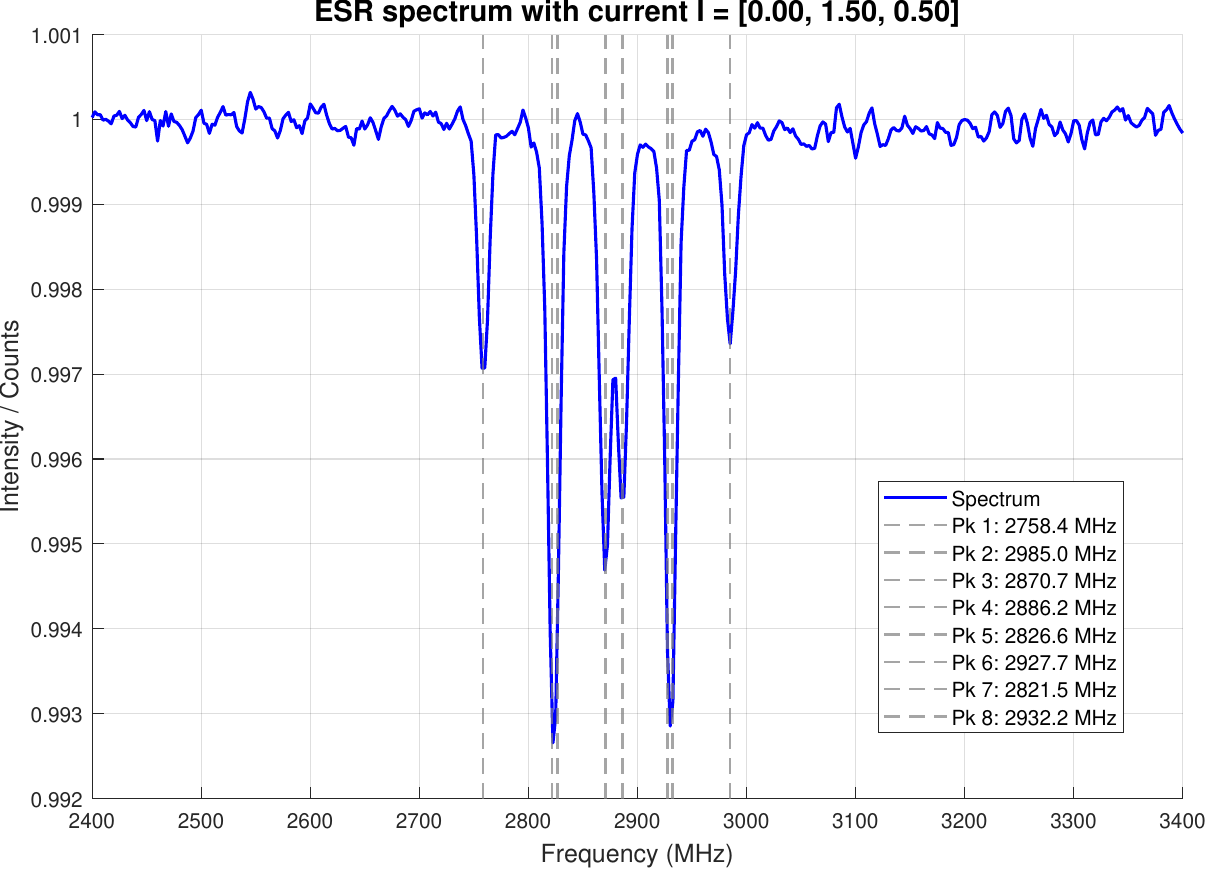}
    \caption{ODMR spectrum with six resolved peaks. The second peak from each side contains two overlapping resonances.}
    \label{fig:demo-six-peaks}
\end{figure}

But for the purpose of this paper, this is not a regime that we can comfortably talking about: Given that the orientations are rather random, we are not guaranteed to have a perfect eight peaks. In that case, the real spectra may only have six or fewer peaks, meaning that our previous verdict is no longer useful. Say in an ideal Zeeman case, where the splittings are proportional to the total magnetic field projected onto the axes, and an NV spectra was measured, where we got six peaks in total, since there are two peaks overlapping with each other. A realistic demo is shown in \ref{fig:demo-six-peaks}

In that case, firstly it would be very hard to tell the differences between these two peaks. To make things even worse, physics constraints also aren't in our sides: The four projected peaks only satisfies the following conditions:

\begin{equation}
    B^{proj, 1} + B^{proj, 2} =  - (B^{proj, 3} + B^{proj, 4})
\end{equation}

While we can explicitly tell the positions of the peaks split by the first magnetic field and the second one with our earlier verdicts, since the third and the forth field are just too close, we cannot tell the peaks between these two, resulting in the following situation of

\begin{equation}
    |{B^{proj, 3} - B^{proj, 4}}| < \delta{B_0}
\end{equation}

where the $\delta{B_0}$ is the limit imposed by the widths of the ODMR spectra. Given that we don't have too much information beyond that for us to truly identify the spectra, the widths of spectra essentially became the noise level we must endure. The same thing would be true if we have even fewer peaks, and the real justification for us to use random noises to simulate the errors in our prediction errors.

\section{Numerical Experiments}

To further quantify the accuracy and robustness (whether there would be spurious results) of our method, we performed numerical simulations mimicking the bulk-diamond experiment (see Supplementary Fig.~\ref{fig:num_exp}). We simulated ODMR spectra using the known axes (Eq.~(5)) and a defined local field $\vec{B}_l = [9, -17, -50]\, \mu{T}$, similar to the measured field. Gaussian noise with a standard deviation of 0.3 MHz (comparable to experimental linewidths for bulk diamond) was added to the simulated resonance frequencies.

The reconstruction results is both accurate in terms of the local fields and consistent such that no spurious results are obtained, as shown in Fig.~\ref{fig:num_exp}. The mean deviation of the predicted local field from the simulation's ground truth was below 10 $\mu{T}$, while the mean angular deviation of the predicted axes from the ground truth was below 0.1$^\circ$. Furthermore, the variance across multiple runs (using different noise instantiations or subsets of simulated bias fields) was low, with the standard deviation of the predicted field components ($\delta B$) below 10 $\mu{T}$ and the standard deviation in axis orientation ($d_{gc}$) below $0.5^\circ$.

\begin{figure}
    \centering
    \includegraphics[width=\linewidth]{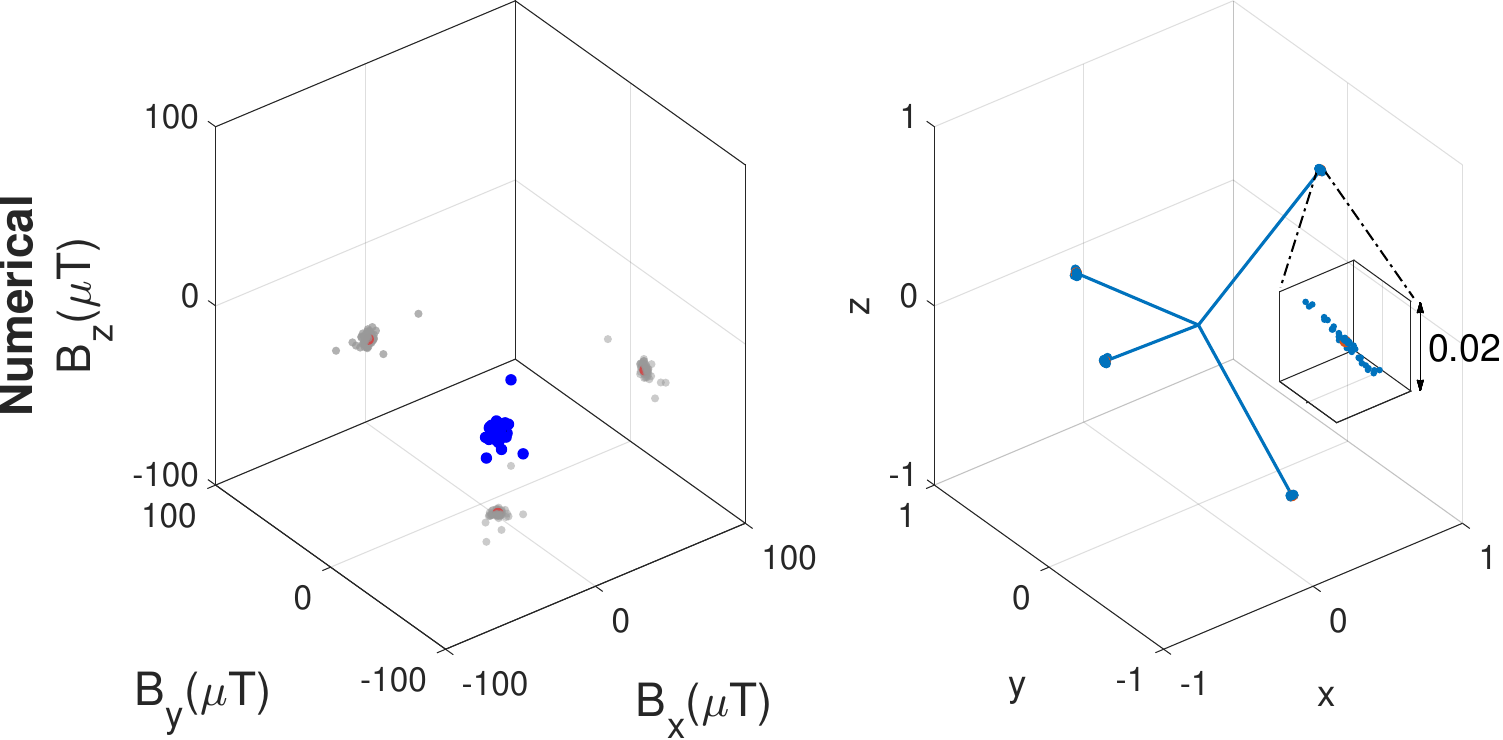}
    \caption{Numerical experiment results for the bulk-diamond setup corresponding to Figs.~3(a) and 3(b) of the main text. Comparable input parameters yield comparable reconstructed fields and axes.}
    \label{fig:num_exp}
\end{figure}

\subsection{Raw Data Examples}
For the isotopically purified bulk diamond used in the proof-of-concept measurement, which has a very low $^{13}\mathrm{C}$ concentration and whose relevant NV centers contain $^{15}\mathrm{N}$ nuclei, the ground-state spin Hamiltonian is:

\begin{equation}
\hat{H} = D_{\mathrm{gs}} S_z^2
+ \gamma_e \vec{B} \cdot \vec{S}
+ \vec{S} \cdot \mathbf{A} \cdot \vec{I}
- \gamma_n \vec{B} \cdot \vec{I}
\end{equation}

where
\( A_\parallel = 3.03~\mathrm{MHz} \) and \( A_\perp = 3.65~\mathrm{MHz} \) \cite{PhysRevB.79.075203}
are the longitudinal and transverse hyperfine coupling constants for the \(^{15}\mathrm{N}\) nuclear spin ($I=1/2$), and
\( \gamma_n = -4.316~\mathrm{MHz/T} \) \cite{doherty2013nitrogen} is the gyromagnetic ratio of the \(^{15}\mathrm{N}\) nucleus. This leads to the hyperfine structure. In bulk-diamond measurements, the narrow linewidth makes this structure resolvable; each of the eight electronic resonances is split into a $^{15}\mathrm{N}$ doublet, giving up to 16 resolved lines. For the analysis shown in Fig.~\ref{fig:raw-data}, we replace each resolved doublet by its midpoint before fitting. Because the $^{15}\mathrm{N}$ nuclear-Zeeman contribution at a field of order $1~\mathrm{mT}$ is only a few kilohertz, the doublet separation can be approximated as $3.03~\mathrm{MHz}$~\cite{PhysRevB.79.075203}.

\subsection{Consistency Checks}

To verify that our methods reproduce the local magnetic fields, we tried to use two different nanodiamonds in the same plate to see whether a consistent local magnetic field can be predicted.

\begin{figure}
    \centering
    \includegraphics[width=\linewidth]{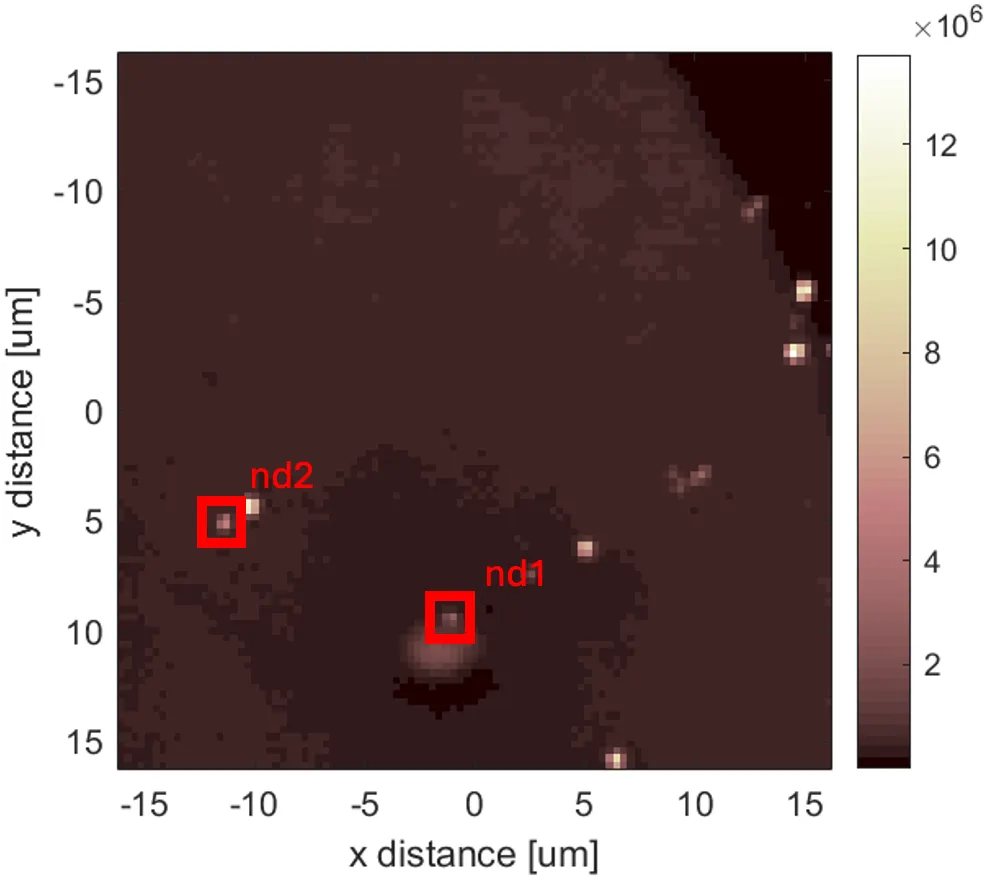}
    \caption{Two nanodiamonds measured on the same substrate.}
    \label{fig:nd-distribution}
\end{figure}

The results are shown below in Fig.~\ref{fig:B-field-caculation}. For nanodiamond 1 (ND1), the predicted local magnetic field is approximately $[11, 10, 46] \pm [11, 14, 27] \, \mu\text{T}$, and for nanodiamond 2 (ND2) it is $[-20, 2, 68] \pm [28, 26, 41] \, \mu\text{T}$. The two results are consistent within the experimental uncertainties.
\begin{figure}
    \centering
    \includegraphics[width=\linewidth]{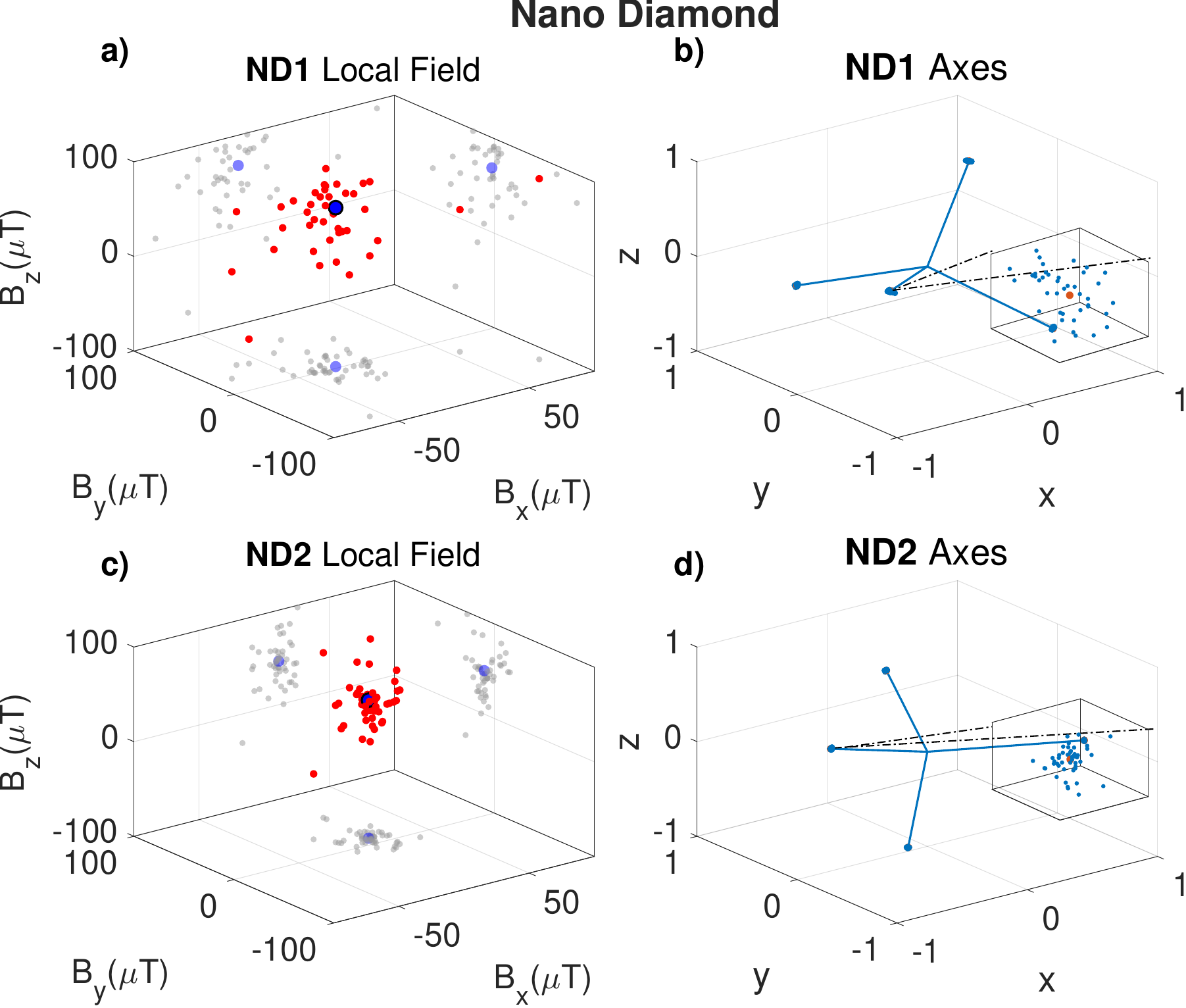}
    \caption{Predicted local magnetic fields of ND1 and ND2 shown in Fig.~\ref{fig:nd-distribution}. The two reconstructed field distributions are consistent within their experimental uncertainties.}
    \label{fig:B-field-caculation}
\end{figure}

\newpage
 \section{Experimental Details}

 \subsection{Setup}

 In this work, we perform optically detectable magnetic resonance (ODMR) spectroscopy to probe the spin-states transitions of the NV under different magnetic environment. Both the laser and microwave are continuous during the measurement, while we sweep the frequency of the microwave signal. When the NV spins and microwave frequency are off-resonant, all population stays in the $\ket{m_s=0}$ state. Once they are in-resonant, the spin population is driven into $\ket{m_s=\pm 1}$, resulting in a decrease in NV fluorescence.

\begin{figure}[htbp!]
    \centering
    \includegraphics[width=0.5\linewidth]{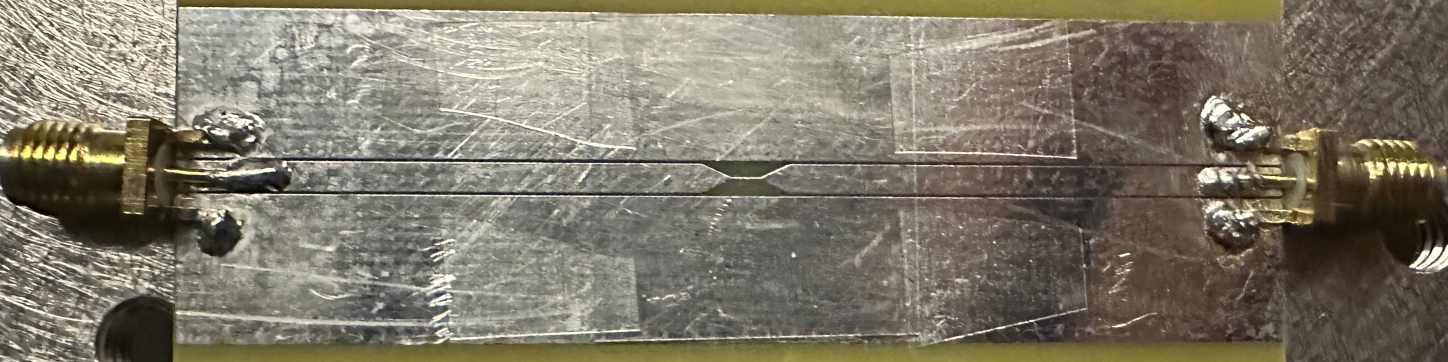}
    \caption{The coplanar waveguide used in experiments}
    \label{fig:waveguide}
\end{figure}

\begin{figure*}
    \centering
    \begin{overpic}[width=0.9\linewidth]{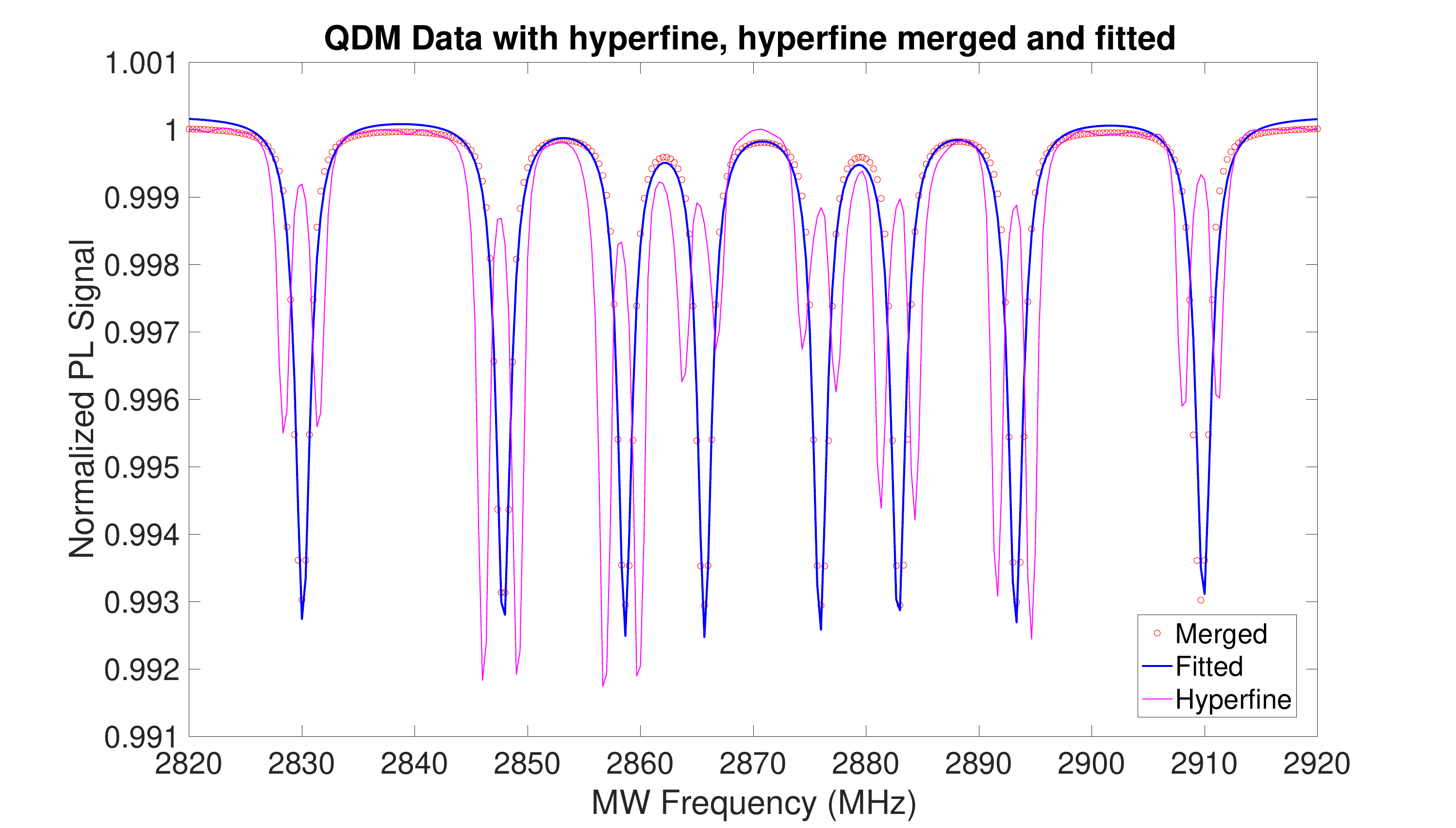}
        \put(50,54){\makebox(0,0){\colorbox{white}{\footnotesize \,\textbf{bulk-diamond data}\textbf{ with hyperfine, hyperfine merged and fitted}}}}
    \end{overpic}
    \caption{Due to bulk diamond's narrow line width, hyperfine peaks are also detectable, while not easily fittable. Therefore, we merge the hyperfine peak by finding the midpoint of the each pair of the hyperfine peaks. The magenta curve shows the raw, unmerged hyperfine data. The red circles are the merged data, and the blue curve is the fitting result}
    \label{fig:raw-data}
\end{figure*}

We perform the nanodiamond measurement on a custom-built confocal scanning microscope. We illuminate the nanodiamonds with a continuous-wave 532 nm laser (Millennia eV, Spectra-Physics). The laser beam is focused onto the sample through a Mitutoyo 100$\times$ M Plan APO objective (NA 0.70, 6 mm working distance, 2 mm focal length), and the fluorescence is collected by the same objective and a single photon counting module (SPCM-AQRH-61-FC,
Excelitas Technologies) with a 633~nm long-pass filter. Microwave control tones for ODMR come from a Stanford Research Systems SG384 signal generator, amplified by an RF amplifier (ZHL-15W-422-S+, Mini-Circuits). The microwave signal is delivered with a custom-made coplanar waveguide, shown in Fig.~\ref{fig:waveguide}. The sensing elements are single-crystal nanodiamonds ($\sim$0.75~$\mu$m, ADAMAS, MDNV1umHi10mg). Static bias fields are provided by coils (Buckley Systems, serial 153 906). The microwave power used for nanodiamond ODMR measurements was typically around 20--25~dBm. This value and the 26--28~dBm value quoted in the referee response refer to different reference planes in the source--amplifier chain (source/input versus amplified output), rather than measurements at a common plane.

\subsection{Sample Preparation}

The nanodiamond sample was prepared using single-crystal NV diamond particles ($\sim$1~$\mu$m diameter, ADAMAS MDNV1umHi10mg) dispersed in rapidly evaporating organic solvents (acetone and methanol). The volume ratio of the ADAMAS stock solution to solvent ranged from 1:50 to 1:100. A total of 0.3--0.4~mL of the mixed solution was deposited onto a cover-glass substrate and spin-coated at 3000~rpm for 60~s. The volume ratio did not affect the ODMR results and was adjusted primarily to facilitate locating individual nanodiamonds on the substrate.

\subsection{Measurement}

We apply magnetic fields by adjusting the current in electromagnets. We use the coil currents as the experimental control parameters (and as the standard for calibration). Because the coil calibration matrix is experiment-dependent, we report the applied coil current configurations for all runs; the corresponding field vectors are reconstructed from the calibration procedure described above. For the bulk-diamond experiment, the applied current configurations are:

\begin{table}[h]
\centering
\begin{tabular}{r r r r}
\toprule
Index & $I_x$ & $I_y$ & $I_z$ \\
\midrule
1  & -0.5 & -2.0 & -2.0 \\
2  & -1.0 & -2.0 & -2.0 \\
3  & -1.5 & -2.0 & -2.0 \\
4  & -2.0 & -0.5 & -2.0 \\
5  & -2.0 & -1.0 & -2.0 \\
6  & -2.0 & -1.0 & -2.0 \\
7  & -2.0 & -1.5 & -2.0 \\
8  & -2.0 & -2.0 & -0.5 \\
9  & -2.0 & -2.0 & -1.0 \\
10 & -2.0 & -2.0 & -1.0 \\
11 & -2.0 & -2.0 & -1.5 \\
12 & -2.0 & -2.0 & -2.0 \\
13 & 1.0  &  2.0 &  2.0 \\
14 & 1.5  &  2.0 &  2.0 \\
15 & 2.0  &  0.5 &  2.0 \\
16 & 2.0  &  1.0 &  2.0 \\
17 & 2.0  &  1.0 &  2.0 \\
18 & 2.0  &  1.5 &  2.0 \\
19 & 2.0  &  2.0 &  0.5 \\
20 & 2.0  &  2.0 &  1.0 \\
21 & 2.0  &  2.0 &  1.0 \\
22 & 2.0  &  2.0 &  1.5 \\
23 & 2.0  &  2.0 &  2.0 \\
24 & 0.5  &  2.0 &  2.0 \\
\bottomrule
\end{tabular}
\caption{Bulk-diamond current configurations ($I_x$, $I_y$, $I_z$) }
\end{table}

Here we measured some configurations multiple times to ensure that drift is tolerable. In addition to the tables provided here, we have posted the calibration scripts online for reproducibility.

For the nanodiamond experiment, the applied current configurations are:

\begin{table}[h]
\centering
\begin{tabular}{r r r r}
\toprule
Index & $I_x$ & $I_y$ & $I_z$ \\
\midrule
1  & 0   & 1     & 0   \\
2  & 1   & 1     & 0   \\
3  & 2   & 1     & 0   \\
4  & 3   & 1     & 0   \\
5  & 4   & 1     & 0   \\
6  & 0   & -2.5  & -0.5 \\
7  & 0   & -2.5  & -1   \\
8  & 0   & -2.5  & -1.5 \\
9  & 0   & -2.5  & 0    \\
10 & 0   & -2.5  & 0.5  \\
11 & 0   & -2.5  & 1    \\
12 & 0   & -2.5  & 1.5  \\
13 & 0   & -2    & -0.5 \\
14 & 0   & -2    & -1   \\
15 & 0   & -2    & -1.5 \\
16 & 0   & -2    & 0    \\
17 & 0   & 2     & 0    \\
18 & 0   & 2     & 0.5  \\
19 & 0   & 2     & 1    \\
20 & 0   & 2     & 1.5  \\
21 & 0   & 2.5   & 0    \\
22 & 0   & 2.5   & 0.5  \\
23 & 0   & 2.5   & 1    \\
24 & 0   & 2.5   & 1.5  \\
25 & 0   & 0     & 1.5  \\
26 & 0   & 0.5   & 0    \\
27 & 0   & 1.5   & 0    \\
28 & 3   & 0     & 0    \\
29 & 0   & 0     & 1.5  \\
30 & 0   & 0     & 1.5  \\
31 & 0   & 0     & 1.5  \\
32 & 0   & 1     & 0.5  \\
33 & 0   & 1     & 1    \\
34 & 0   & 1     & 1.5  \\
\bottomrule
\end{tabular}
\caption{Nanodiamond current configurations ($I_x$, $I_y$, $I_z$) }
\end{table}

\bibliography{apssamp}